\documentclass[10pt]{article}

\usepackage{amsfonts}
\def\form#1{(\ref{#1})}
\def\Co{I \kern-.66em C}

\def\c{\mathcal}

\def\al{\alpha}

\def\pa{\partial}

\def\c#1{{\cal{#1}}  }
\def\L{{\cal L}}               
  \def\B{{\cal B}}
\def\H{{\mathcal H}}
\def\Q{{\mathcal Q}}              

\def\ram{\mathop{\longrightarrow}\limits}

\def\dim{\mathop{\rm dim}\nolimits}
\def\Tr{\mathop{\rm Tr}\nolimits}

\def\CS{\mathop{\rm CS}\nolimits}
\def\CCS{\mathop{\rm CScov}\nolimits}

\def\Aut{\mathop{\rm Aut}\nolimits}
\def\Gauge{\mathop{\rm Gauge}\nolimits}
\def\Diff{\mathop{\rm Diff}\nolimits}

\renewcommand{\Re}{I\kern-.36em R}         

\newcommand{\be}{\begin{equation}}
\newcommand{\ee}{\end{equation}}
\newcommand{\ba}{\begin{eqnarray}}
\newcommand{\ea}{\end{eqnarray}}
\newcommand{\baa}{\be\left\{\begin{array}{l}}
\newcommand{\eaa}{\end{array}\right.\ee}

\def\QDE{\rule{2.5mm}{2.5mm}}
\def\CVD{$\phantom{'}$\hfill\QDE}

\newtheorem{Theorem}{Theorem}[section]

\newtheorem{Remark}[Theorem]{Remark} %
\newtheorem{Definition}[Theorem]{Definition} %
\newtheorem{Lemma}[Theorem]{Lemma} %
\newtheorem{Example}[Theorem]{Example} %
\newtheorem{Proposition}[Theorem]{Proposition} %
\newtheorem{Exercise}[Theorem]{Exercise}%

\title{Covariant Charges \\ in Chern--Simons   $\hbox{AdS}_3$ Gravity}
\author{G.\ Allemandi\thanks{E-mail:
allemandi@dm.unito.it},
M.\ Francaviglia\thanks{E-mail:
francaviglia@dm.unito.it}, M.\ Raiteri\thanks{E-mail:
raiteri@dm.unito.it}
\\
Dipartimento di Matematica, Universit\`a degli
Studi di Torino,\\
Via Carlo Alberto 10, 10123 Torino, Italy }

\date{}

\begin{document}
\maketitle

\begin{abstract}
We try to give hereafter an answer to some open questions about the 
definition of
conserved quantities in Chern-Simons theory, with particular reference
to Chern-Simons $\hbox{AdS}_3$ Gravity. Our attention is focused on  the
problem of global covariance and gauge invariance of the variation of
Noether charges.
A theory which satisfies the principle of covariance on each step of its
construction is developed, starting from a gauge invariant
Chern--Simons Lagrangian   and using a recipe developed
in 
\cite{HIO} and
\cite{forth}  to calculate the variation of conserved  quantities.
The problem to give a mathematical well--defined expression  for the
infinitesimal generators of symmetries  is pointed out and it is shown
that  the generalized Kosmann lift of spacetime vector fields  leads to  
the expected numerical values for the conserved quantities when  the
solution corresponds to the BTZ black hole. The fist law  of black holes
mechanics for the BTZ solution  is then  proved and the transition between
the variation of conserved quantities in Chern-Simons
$\hbox{AdS}_3$ Gravity   theory and the variation of conserved quantities
in General Relativity is analysed in detail.\\

\centerline{\it Dedicated to \emph{Matteo Raiteri}, born in the Fifth of
October 2002 }

\end{abstract}

\section{Introduction}

Chern-Simons field theories 
have been widely studied in the past decades as a possible  model to analyse
the classical and quantum behaviour  of the gravitational field. Efforts
were focused towards  rewriting gravity as a gauge theory  with gauge group
the Poincar\'e group or the (anti) de Sitter group.
 To this purpose, in place of the Hilbert-Einstein Lagrangian, a
Chern-Simons Lagrangian was considered in
which the gauge potential  is  a linear combination of the vielbein
and the  spin connection. This is possible in all odd dimensions and
particularly in dimension
three, where field equations reproduce exactly the Einstein field
equations; see \cite{Ach,BH3,BH4,Ca,Wi1} where  it is
shown  that
$2+1$ gravity with a negative cosmological constant can be formulated as
a  Chern-Simons theory (see also \cite{CF} for higher dimensional
Chern--Simons gravity). In particular it was  found in \cite{Mann93} that a
particular solution of Chern-Simons theory corresponds to the well--known
BTZ black hole
\cite{btz}.\\

The interest in  $3$-dimensional Chern-Simons theory as a possible and
simpler model to analyse $(2+1)$-dimensional gravity   was also
strengthened by the observation that the thermodynamics of higher
dimensional black  holes can be understood in terms of the BTZ
solution.   The BTZ solution   provides indeed a model for the
geometry of a great amount of black hole solutions relevant to string
theory, the geometry of which can be factorized in
the  product of  $BTZ \times M$, where $M$ is a simple manifold; see
\cite{Hyun}.
\\

According to the renewed interest in Chern--Simons theories 
a lot of papers  dealing with the
problem of gauge symmetries and gauge charges for Chern-Simons theories
have appeared in the recent years, all  addressed  to analyse
the origin of the gravitational boundary degrees of freedom and,
eventually,  to understand  the statistical mechanical origin of
Bekenstein-Hawking entropy via a micro-- and grand--canonical calculation
(see \cite{BH3,BH4,BH,Ca} and references therein). \\

Motivated by  this state of affairs, in  the present  article we deal with
the problem of   developing   a totally covariant  formulation for    
Chern-Simons conserved quantities, with particular attention to
$SL(2,\Re)\times SL(2,\Re)\simeq SO(2,2)$ Chern--Simons theory in dimension
three (with this gauge choice, Chern--Simons theory is well suited to
describe AdS gravity; see e.g
\cite{Ach,BH3,BH4,Mann93,Ca,Wi1}). This issue is tackled, first of all,  
by framing Chern--Simons theory   into  the mathematical domain of gauge
natural theories, which provide a unifying mathematical arena to describe
all classical Lagrangian field theories and, in particular, are
fundamental to mathematically   analyse the interaction of  gravity with
fermionic and bosonic matter; see e.g.
\cite{Lorenzo,Lorenzo2}.  The gauge natural  approach is
essentially based on the Lagrangian formulation of field
theories in terms of fiber bundles and the Calculus of
Variations on jet bundles. Hence, in order to have a gauge
natural formulation of Chern--Simons theory, the first
problem we meet   is to construct a covariant Chern--Simons
Lagrangian, where covariance is referred both to  spacetime
diffeomorphisms and gauge  transformations.  Indeed, despite
the field equations are covariant for  both spacetime
diffeomorphisms and  gauge transformations,
 the Chern-Simons Lagrangian is not covariant  for
gauge transformations. 
To solve this problem we define a covariant Lagrangian for the theory 
which differs from the usual Chern-Simons Lagrangian  just for the
addition of  a divergence term. In such a way a \emph{covariantized}
Lagrangian is obtained and, in the meantime,  the field equations remain
unchanged.  The Lagrangian  we calculate  hereafter 
 is the same ``covariantized''  Lagrangian obtained in
\cite{BFF01}   through the
use of the ``transgression'' formula \cite{CH1}.
In the same paper \cite{BFF01} a calculation of conserved quantities based
on Noether  theorems and  the covariant formula for the superpotential of
the theory were obtained. Nevertheless, we shall  show in the present paper
that the superpotential, even though it is a well--defined mathematical
object, in practice it is not suited to calculate conserved
quantities. Indeed when conserved quantities are explicitly
calculated for the Chern--Simons solution corresponding  to the BTZ
black hole  via the integration of the superpotential on a  circle
enclosing the horizon,  the expected values for mass
and angular momentum are not recovered even if the circle tends to spatial
infinity (see equation \form{202} below). This is a rather undesiderable
result. In fact, even if the same physical solution  of a field theory  can
be obtained  from two, or more,  different field theories (whose field
content  is nevertheless equivalent on--shell)  it would be reasonable 
that physical observables  depend  solely on the given solution, e.g. the
black hole,  and  not  on  the theory the solution  comes from.  Roughly
speaking,  mass, angular momentum and charge  of a black hole are physical
properties of the black hole itself  and they should  not change  if the
same black hole solution is obtained either from the Hilbert Lagrangian 
of General Relativity  or from the Chern--Simons Lagrangian. Hence, the
Noether charges  which generate  conserved quantities  have to be somehow
tied to the solution  under examination  rather than to the
Lagrangian which generates the equations of motion.
Since the Chern--Simons Lagrangian can be mapped into the Hilbert
Lagrangian only modulo divergence terms, 
 this basically means that   the general formula
for  the conserved  currents  we are interested 
in has to be linked only to the homology class of
the Lagrangian (i.e., it has not to depend on additional divergence terms).
In this way conserved quantities are related to the equations of motion
and  do not depend on the representative chosen  inside the
homology class of Lagrangians.

A step towards the
solution of this problem can be found in  the papers
\cite{Silva}, whereby conserved quantities  are obtained from the
equations of motion via a cascade equation.
Similarly, in the present paper   we shall  mainly be concerned with
the issue of developing  a fully covariant  approach to conserved
quantities in Chern--Simons gravity which  could lead to   a
direct correspondence with charges in General Relativity, thus
giving the expected  values for mass, angular momentum and entropy for
the BTZ solution. \\
The main problems to be solved   are  two.
The first one deals with the very basic object which has to enter in
the definition of conserved quantities. Indeed, we will check
  that the superpotential does not reproduce the correct
numerical values,
even if  in a first approach to
conserved quantities  it seems to be the best candidate  to describe
global charges.
The reason of these discrepancies  is mainly due  to the fact that
the naive definition of charges via the superpotential alone does not
take into a proper account  the role played  by boundary terms.
Boundary conditions  have to be imposed  on the dynamical fields,
namely Dirichlet or Neumann conditions, specifying which are the
intensive or extensive variables or,  equivalently, which are the
control--response parameters. The choice of boundary conditions
basically reflects  onto the choice of a background solution the
starting solution has to be matched with. Physically speaking, the
background fixes the zero level for all conserved quantities
(\cite{Booth,BY,Nester,Remarks,CADM}). A way to overcome  the problems
which come from the background fixing procedure  is to define \emph{the
variation}  of conserved quantities, as suggested in \cite{Nester,CADM,
forth,Silva}. In this way a covariant analysis of boundary terms,
\emph{\'a la} Regge--Teitelboim \cite{RT},   can be
implemented,   leading to the \emph{covariant ADM} formalism for
conserved quantities. This is exactly the
approach we shall develop also for Chern--Simons theory.

The second problem we are faced with is related to  the  choice
of the symmetry generators. This mathematical problem  is by no means 
trivial  and it deserves a careful  investigation. Indeed,
in a natural theory such as General Relativity the action of spacetime
vector fields on the dynamical fields is unambiguously defined. This means
that we know how the fields are Lie dragged along spacetime directions and
this is enough to build a mathematical consistent theory of conserved
quantities. On the contrary,  in a theory with gauge invariance (such
as  Chern--Simons theory  as well as
any gauge natural theory; see
\cite{Kolar}) there is no  canonical way  to construct conserved
quantities starting from  a given spacetime vector field.  Indeed, in those
theories there is no canonical way  to  lift the spacetime 
diffeomorphisms  on the configuration space  of the theory, i.e.  the
target space  where fields take their values. On the contrary, there may
exist  different  non canonical (but nevertheless global and hence
well--defined)  ways  to perform lifts of spacetime diffeomorphisms. This
in turn implies  that the transformation rules of the dynamical fields
under spacetime diffeomorphisms are not  defined until  a preferred lift
procedure has been somehow selected. For example we shall see that  in the
 Chern--Simons theory 
there are exactly  three distinct dynamical connections. All of them can  be
equally well used, from a mathematical viewpoint, to define lifted  
vector
fields on the configuration bundle starting from spacetime
diffeomorphisms. These vector fields, in their turn,  enter into the
definition of Lie derivatives  of the dynamical fields  and they
eventually lead  to different definitions of conserved quantities.
All these  definitions provide well--defined and mathemathical
meaningful expressions. But which of them is also physically
meaningful? In this paper we have tried   to answer this question
by testing all the admissible definitions of conserved quantities
with  the BTZ black hole solution. The numerical results obtained suggest 
that the only viable definition of (the variation of) conserved quantities
inside the Chern--Simons gravity  framework is the one based on the
generalized Kosmann  lift of spacetime diffeomorphisms
\cite{Lorenzo,Godina}. Conserved quantities computed with the generalized
Kosmann lift   reproduce,  in fact, exactly  the expected values for the
BTZ mass, angular momentum and entropy, while other choices do not lead to
meaningful results. Moreover, when we perform the transition from
Chern--Simons
$\hbox{AdS}_3$ gravity to General Relativity,  the formula
expressing the (variation of the) conserved quantities in Chern--Simons
gravity is  mapped   exactly  into  the formula for (the variation  of)
the  conserved quantities in  General Relativity  found in
\cite{Booth,BY,forth,Kij}, as one should expect.\\

\noindent 
The present paper is organized as follows.
In 
Section 2 we define  the covariant Lagrangian for  $SL(2,\Re)\times
SL(2,\Re)$  Chern--Simons theory.
In Section 3 we illustrate the  geometric framework  for the  Hamiltonian
and symplectic formulation of the theory and  we derive a general
formula to calculate the variation of conserved  quantities. In 
Section 4 we 
analyse the problem of defining  the lift to the
configuration bundle for  an infinitesimal generator of symmetries over
spacetime and we define  the generalized Kosmann lift. In Section 5 we 
explicitly calculate the transition of the variation of conserved
quantities from Chern-Simons theory to
$(2+1)$--dimensional gravity. In the  Appendix A are summarized the
notations  and the formulae  entering the calculations of Section 5.
In Appendix B the formalism developed throughout the paper  is
applied to the general anti--de Sitter solution and to the one--particle 
solution of Chern--Simons theory.


\section{The Covariant Chern-Simons Lagrangian}

The $3$--dimensional Chern--Simons Lagrangian   can be written
as:
\be
L_{\CS}(A)= {\kappa\over 4\pi}\epsilon^{\mu\nu\rho}\Tr\left(
A_\mu d_\nu A_\rho+{2\over 3} A_\mu\,A_\nu\, A_\rho\right)\, d^3x
\label{LCS}
\ee
where $\kappa$ is a constant which will be fixed later,  while
$A_\mu=A_\mu^i\, J_i$ are the coefficients of the connection
$1$--form $A=A_\mu\, dx^\mu$ taking their values in any Lie algebra
$\mathfrak g$ with generators
$J_i$. By  fixing  $\mathfrak g=sl(2,\Re)$ and choosing the
generators
\be
J_0={1\over 2}\left(\begin{array}{ccc}
0&-1\\
1&0
\end{array}\right),
\quad J_1=
{1\over 2}\left(\begin{array}{ccc}
1&0\\
0&-1
\end{array}\right),
\quad J_2={1\over 2}\left(\begin{array}{ccc}
0&1\\
1&0
\end{array}\right)\label{base}
\ee
we have  $[J_i,J_j]=\eta^{lk} \epsilon_{kij} \,J_l$  and $\Tr(J_i
J_j)=1/2\, \eta_{ij}$, with $\eta={\mathop{\rm
diag}\nolimits}(-1,1,1)$ and $\epsilon_{012}=1$. Hence, the Lagrangian
\form{LCS} can be explicitly written as:
\be
\begin {array}{rl}
L_{\CS}(A)&= {\kappa\over 8\pi}\epsilon^{\mu\nu\rho}\left(
\eta_{ij} A^i_\mu d_\nu A^j_\rho+{1\over 3}\epsilon_{ijk}
A^i_\mu\,A^j_\nu\, A^k_\rho\right)\, d^3x\\
\\
&= {\kappa\over 16\pi}\epsilon^{\mu\nu\rho}\left(
\eta_{ij} F^i_{\mu \nu} A^j_\rho-{1\over 3}\epsilon_{ijk}
A^i_\mu\,A^j_\nu\, A^k_\rho\right)\, d^3x
\end{array}\label{LSL}
\ee
where $ F^i_{\mu \nu}=d_\mu A^i_\nu-d_\nu A^i_\mu+ \epsilon^i_{jk}
A^j_\mu A^k_\nu$ is the field strength.

We then consider two  independent  $sl(2,\Re)$
 dynamical
connections $A$ and 
$\bar A$,  the evolution of which
is dictated by the  Lagrangian
\be
L_{\CS}(A,\bar A)=L_{\CS}(A)
-L_{\CS}(\bar A)\label{Ldoppia}
\ee
which is nothing but the difference  of two 
Chern--Simons Lagrangians \form{LSL}, one for each dynamical
connection. Field equations ensuing from \form{Ldoppia} are of course:
\be
\left\{
\begin{array}{l}
\eta_{ij}\, \epsilon^{\mu\nu\rho}\,
F^i_{\mu\nu}=0\\
\\
\eta_{ij}\, \epsilon^{\mu\nu\rho}\, \bar F^i_{\mu\nu}=0
\end{array}
\right.\label{fieldequ}
\ee
Starting from the fields $A,\bar A$ it is then possible (see
\cite{Ach,BH3,Mann93,Wi1}) to define two new dynamical fields, $e^i$ and
$\omega^i$, through the rule:
\be
A^i= \omega^i +{1\over l} \, e^i\qquad
\bar A^i= \omega^i -{1\over l} \, e^i\quad (l=\hbox{constant})\label{trans}
\ee
In terms of the new $(e, w)$ variables field equations \form{fieldequ}
become:
\be
R^{ij}=-{1\over l^2} e^i\wedge e^j ,\qquad
T^i=de^i+ \omega^i_j \wedge e^j=0\label{Einsteineq}
\ee
where
$\omega^i=1/2
\,
\eta^{ij}\,
\epsilon_{jkl}
\, \omega^{kl}$ and
$R^i_j =d\omega^i_j + \omega^i_k \wedge \omega^k_j$.
Equations \form{Einsteineq} are nothing but Einstein's equations with
cosmological constant $\Lambda=-1/l^2$ written in terms of the triad
field
$e^i$ and the torsion--free spin connection
$\omega^i_j$. Moreover
the Lagrangian
\form{Ldoppia}, in the new variables, reads as:
\be
\begin{array}{rl}
L_{\CS}(A(w,e),\bar A(w,e))&={\kappa\over 4\pi\,l}\sqrt{g}  (
g^{\mu\nu}\, R_{\mu\nu}- 2\Lambda)+d_\mu\left\{ {k\over
8\pi}\eta_{ij}\epsilon^{\mu\nu\rho} A^i_\nu \,
\bar A^j_\rho
\right\}\\
\\
&={\kappa\over 4\pi\,l}\sqrt{g}  (
g^{\mu\nu}\, R_{\mu\nu}- 2\Lambda)+d_\mu\left\{ {k\over
4\pi l}\eta_{ij}\epsilon^{\mu\nu\rho} e^i_\nu \,
\omega^j_\rho
\right \} \label{daCSaG}
\end{array}
\ee
with  $g_{\mu\nu}=\eta_{ij}\, e^i_\mu\, e^j_\nu$ and $R_{\mu\nu}=
R^i_{j\rho\nu}\, e^j_\mu\, e^\rho_i$  being the
Ricci tensor of the metric $g$.
Notice, however, that the transition from Chern--Simons theory to General
Relativity displays some theoretical undesiderable features.
Indeed, while Chern--Simons equations of motion are
manifestly covariant with respect to spacetime diffeomorphism as well
as with respect to gauge transformations, the  Chern--Simons 
Lagrangian
\form{LSL} is not gauge invariant. If we consider a pure gauge
transformation with generator $\xi^i$ acting on the gauge potential,
i.e.
\be
\delta_\xi \,A^i_\mu=D_\mu \xi^i
\ee
from \form{LSL} we obtain:
\be
\delta_\xi L_{\CS}(A)=d_\mu\left\{{\kappa\over
8\pi}\epsilon^{\mu\nu\rho}\, \eta_{ij} A^i_\nu\, d_\rho \xi^j\right\}
\ee
The fact that \form{Ldoppia} is not gauge invariant   becomes even more
explicit in expression
\form{daCSaG} where the non invariant term is ruled out  into the
divergence part.  The simplest
way to overcome this drawback is to push the divergence term appearing in
the right hand side of
\form{daCSaG} into the left hand side, defining in  this way a global
covariant Chern--Simons Lagrangian $L_{\CCS}$:
\ba
L_{\CCS}(A,\bar A)
&=&{\kappa\over 8\pi}\epsilon^{\mu\nu\rho}\left(
\eta_{ij} A^i_\mu d_\nu A^j_\rho+{1\over
3}\epsilon_{ijk} A^i_\mu\,A^j_\nu\,
A^k_\rho\right)\, d^3x\nonumber\\
&-&{\kappa\over 8\pi}\epsilon^{\mu\nu\rho}\left(
\eta_{ij} \bar A^i_\mu d_\nu \bar A^j_\rho+{1\over
3}\epsilon_{ijk} \bar A^i_\mu\,\bar A^j_\nu\,
\bar A^k_\rho\right)\, d^3x\label{11}\\
&-&d_\mu\left\{
{k\over 8\pi}\eta_{ij}\epsilon^{\mu\nu\rho} A^i_\nu \,
\bar A^j_\rho\right\}d^3x\nonumber
\ea
or, equivalently:
\be
L_{\CCS}(A,\bar A)={k\over
8\pi}\epsilon^{\mu\nu\rho}\left\{\eta_{ij}\, \bar F^i_{\mu\nu}\,
B^j_\rho + \eta_{ij}\,\bar D_\mu B^i_\nu\, B^j_\rho +{1\over 3}
\epsilon_{ijk} B^i_\mu\, B^j_\nu\, B^k_\rho\right\}d^3x\label{LB}
\ee
\\
where $\bar D_\mu$ is the covariant derivative with respect to the 
connection $\bar A$ and we set $B^i_\mu=A^i_\mu-\bar A^i_\mu$.
Being  $B^i_\mu$  tensorial, expression
\form{LB} tranforms as a scalar function under gauge transformations
(and  as a scalar density under diffeomorphisms).
Taking
into account  definition
\form{trans} we now obtain:
\be
L_{\CCS}(A(\omega,e),\bar A(\omega,e))={\kappa\over 4\pi\,l}\sqrt{g} \,(
g^{\mu\nu}\, R_{\mu\nu}- 2\Lambda)
\ee
meaning that  the Chern--Simons Lagrangian \form{11}  is mapped 
exactly (i.e. without undesiderable non invariant  boundary terms) into
the Hilbert Lagrangian for General Relativity with a negative cosmological
constant, provided   we
set
\be
\kappa={l\over 4G}
\ee
being $G$ the Newton's constant (and setting $c=1$).
We remark that the Lagrangian \form{11},  or equivalently \form{LB}, is the
same ``covariantized''  Lagrangian already obtained in \cite{BFF01} 
through the
use of the ``transgression'' formula established by Chern and Simons
(see \cite{CH1}). It was  shown in \cite{BFF01} that a  ``covariantization''
 procedure can be applied to each Chern--Simons Lagrangian in
dimension three, independently on the relevant gauge group of the
theory.


\section{Variation of Noether Charges}

\label{Nother Theorem}
  The approach here proposed to calculate the fundamental  parameters of 
the theory, such as mass, angular momentum and  gauge charges, is  
based on a geometrical Lagrangian formalism for classical field 
theories. In this framework conserved currents and conserved quantities 
can be calculated by means of the first and the second Noether theorem 
as shown in  \cite{BFF}, \cite{Lagrange}, \cite{forth}. The Hamiltonian 
formalism for the theory can also be derived identifying 
the variation of the Hamiltonian  with
the variation of the  Noether current with respect to a  
vector field transversal to a Cauchy hypersurface in spacetime 
\cite{forth}. The variation of energy is then naturally defined as
the on-shell value of the variation of the Hamiltonian. The advantages
which derive in using this approach are  related to the fact that 
all  quantities we are going to introduce (Noether currents,
 Noether  charges and symplectic forms) are both covariant and gauge
invariant, thereby  having  a global  geometrical interpretation.
Physically speaking this means that all formulae retain their validity
independently on the observer, i.e. independently on the coordinate
system in which formulae can be expressed and independently on the
spacetime splitting into space
$+$ time. The whole theory is  also  independent on the addition of
divergence terms to the Lagrangian. Hence we have not to care about
choosing a representative inside the cohomology class of the
Lagrangians. We only have to care that the representative Lagrangian 
be covariant in order to frame the whole theory of conserved
quantities in a well--posed geometric background (this is the
ultimate reason why we have chosen the Lagrangian
\form{11}  in place of  \form{Ldoppia}: the latter Lagrangian
is not gauge  invariant!).

Moreover we shall see that in presence of Killing vector fields   the
variation of the Noether charges,  which are naively defined through
integration on a
$(n-2)$-dimensional surface $B$ in spacetime,
does not change inside the 
homology class of 
$2$-dimensional surfaces to which $B$ belongs  (see \cite{forth}).
Roughly speaking this is the mathematical property which will allow
to formulate  the first law of black holes mechanics. 

In addition,  in the framework we shall develop  we do not have  to
impose \emph{a priori}   boundary conditions to make the variational 
principle well defined.  Boundary conditions just assume  a
fundamental  role, \emph{a posteriori},  in the formal integration of
the variational equation  which defines the  Noether charges, e.g. the
variation of energy.  Different boundary conditions on the fields
(corresponding to different ways in which  the  physical system  
can interact with the outside)
can be
imposed on the same variational equation  leading to different
physical  interpretations of the results, e.g. internal energy for
Dirichlet boundary conditions, free energy  for Neumann boundary
conditions and so on; see \cite{HIO,BY,Nester,forth,Kij}\\

We shall assume that the reader is already familiar with the
geometrical language of fiber bundles and with the calculus of
variations on jet bundles (see, e.g. \cite{Lagrange,Kolar,Saunders}). We
just recall  few notions in order to fix the notation.
 As it is common use in a geometric approach to field theories  
  the Lagrangian
$L(j^k
\varphi)=\L(j^k \varphi)\, ds$ is considered as  a global horizontal 
$m$--form  on the $k$-order prolongation of the configuration
bundle $Y\ram M$ (which is a fiber bundle over the base manifold
$M$, with $\dim M=m$),
$\L$ is the Lagrangian density,
$ds$  is the volume form on $M$ (in a coordinate
chart  $ds=dx^1\wedge\dots \wedge dx^m$)  while 
$\varphi$ are
 the fields of the theory, considered as  sections of the
configuration bundle, i.e. $\varphi: M\ram Y$. In the sequel we shall
be mainly concerned with first order theories, i.e. theories in which
the Lagrangian depends only on the fields together with  their first
derivatives  (for  higher order theories we refer the interested reader to
\cite{forth}). As it is well known, the  variation of
the  Lagrangian, after integration by parts,  splits into the sum
of two  terms,  called, respectively, the Euler-Lagrange and the
Poincar\'e-Cartan  morphisms:
\begin{eqnarray}
\delta_X L (j^1\varphi)=<\Bbb{E} (L, j^2 \varphi),
X>+  d  \; <\Bbb{F} (L, j^1\varphi), X>
\label{dL}
\end{eqnarray}
where $X$ is any  vertical vector field on the configuration bundle
(namely, $X=\delta \varphi {\partial\over\partial \varphi}$ describes
the infinitesimal deformation of the dynamical fields),  while
$<\,, \,>$ denotes the canonical pairing between differential forms
and vector fields. Locally:
\be
\begin{array}{rcl}
<\Bbb{E} (L, j^2 \varphi), X>&=&\left\{{\partial \L\over\partial 
\varphi^A}-d_\mu {\partial \L\over\partial (d_\mu
\varphi^A)}\right\} X^A\, ds 
\\
\\
<\Bbb{F} (L, j^1\varphi), X>&=&{\partial \L\over\partial( d_\mu
\varphi^A)} \, X^A\, ds_\mu \qquad (ds_\mu=\partial_\mu \rfloor ds)
\end{array}\label{166}
\ee 
where we have collectively  labelled the fields  with the index $A=1,\dots,
n=\dim Y -\dim M$.
The Euler-Lagrangian morphism  selects the critical
sections
$\varphi$ (i.e. the physical field  solutions) through the field equations
$\Bbb{E} (L, j^2 \varphi)=0$. 

According to \cite{Lagrange,Trautman}, for any given  projectable 
vector field 
$\Xi$ on the manifold $Y$ projecting onto the vector field $\xi$ on
$M$ and locally described by: 
\be
\Xi=\xi^\mu
\partial_\mu+ X^A \partial/\partial \varphi^A,\qquad \xi=\xi^\mu
\partial_\mu\label{88}
\ee
 we say that 
the Lagrangian $L$ admits a
$1$--parameter group of symmetries generated by the 
vector field 
$\Xi$  if it satisfies the following  property:
\begin{eqnarray}
\delta_\Xi L  (j^1\varphi): =  <\delta L,   j^1 \pounds_{{\Xi}}
\varphi > =\pounds_{\xi} L 
\label{simmetr}
\end{eqnarray}
 where $\pounds_{{\Xi}}
\varphi=T\varphi\circ  \xi -\Xi\circ \varphi$ is the geometrically 
defined Lie derivative
of the section $\varphi$ with respect to $\Xi$ and  $\pounds_{\xi}$
denotes the usual Lie derivative of differential forms; see
\cite{Kolar,Saunders}.
 From  equations
(\ref{dL}) and  (\ref{simmetr})  it follows  that the Noether
currents generated by the  infinitesimal symmetry
$\Xi$ can be written as:
\be
\c{E}  ( L, \varphi, \Xi)=<\Bbb{F} ( L, j^1\varphi),
\pounds_{{\Xi}}
\varphi> - i_\xi {L} (j^1\varphi) \label{corr}
\ee
and satisfy  the conservation law $d \c{E}  ( L, \varphi, \Xi)=-
<\Bbb{E} (L, j^2 \varphi),
\pounds_{{\Xi}}
\varphi>$.
 The Noether current
(\ref{corr}) is then a
$(m-1)$-form closed on--shell which can be integrated   on any  
hypersurface
$\Sigma$ of spacetime $M$. 

The field theories we shall  be  concerned with  from now on  are 
\emph{the gauge natural theories} (see
\cite{Remarks,Lorenzo,Godina,Kolar}). In gauge natural theories  the
configuration bundle $Y$ is associated to a given  principal bundle
$P\ram^G M$, with Lie group $G$. Moreover   each projectable vector field
on
$P$, which can be locally written as
$\Xi_P=\xi^\mu(x)\partial_\mu+\xi^i(x,g)\,\rho_i$ (having denoted with
$\rho=(g\,\partial/\partial g)$, $g\in G$, a basis for right
invariant vector field on
$P$ in a trivialization $(x,g)$ of $P$) canonically induces a vector
field
\form{88} on the configuration bundle for which  the property
\form{simmetr} holds true. 
Roughly speaking
gauge natural theories are the ones  which admit  the group $\Aut(P)$
of the automorphisms
 of a principal bundle $P$ as group of symmetries. 
For instance, Yang--Mills theories on a
dynamical background as well as the  covariant Chern-Simons Lagrangian
\form{11} are examples of gauge natural theories. Each vector field  
$\Xi_P=\xi^\mu\partial_\mu+\xi^i\,\rho_i$ on the relevant principal
bundle $P$ induces on the bundle of connections (the sections of which
are the connection $1$--forms  $A^i_\mu$)  the vector field
\be
\Xi= \xi^\mu\partial_\mu+ \Xi^i_\mu {\partial\over \partial 
A^i_\mu}\qquad \Xi^i_\mu=-d_\mu \xi^\nu\, A^i_\nu- C^i_{jk}\,
A^j_\mu\, \xi^k -d_\mu \xi^i\label{20}
\ee
Notice that vertical vector fields, i.\ e.\ the generators of ``pure''
gauge transformations, denoted from now on as $\Gauge(G)$, are the
ones for which
$\xi^\mu=0$; in this case $\Xi^i_\mu=-D_\mu \xi^i $. Each vector
field
\form{20} satisfies the property
\form{simmetr}, thereby inducing a Noether current \form{corr}.

Specifically, in the Chern--Simons theory \form{11}, the relevant principal
bundle $P$ of the theory  is a $SL(2,\Re)$ principal bundle and 
the connections $A$ and $\bar A$  are  two different sections of the
associated configuration bundle $Y=J^1P/SL(2,\Re)$ which is called \emph{the
bundle of connections}; see \cite{Kolar,Saunders}.
 \\

The generalization to gauge natural theories of the
second Noether theorem (see \cite{Remarks,Lagrange,CADM,Robutti}) states
that in each  gauge natural  theory the Noether current can be canonically
split, through an  integration by parts,  into two terms:
\be
\c{E}  ( L, \varphi, \Xi )=\tilde{\c{E}}  ( L, \varphi, \Xi)+ d
[ {U}  ( L, \varphi, \Xi)]  \label{corr1}
  \label{corcon}
\ee
called, respectively,  the \emph{reduced current} $\tilde{\c{E}}$,
which identically vanishes  on shell since it is proportional to
field equations, and the
\emph{superpotential}
${U}
$ of the theory. The reduced current $\tilde{\c{E}}$ is unique while the
superpotential ${U}$ is unique modulo cohomology. It can be uniquely fixed
by choosing a specific connection and integrating covariantly (see
\cite{Robutti}).
\\
\begin{Example}{\rm
The superpotential $U(L_{\CCS},A,\bar A, \Xi)= 1/2\, U^{\mu\nu}\,
ds_{\mu\nu}$ written in local coordinates (with 
$ds_{\mu\nu} = \partial_\nu\rfloor \partial_\mu\rfloor ds$), associated
to the Lagrangian \form{LB} and relative to the 
vector field \form{20} has been calculated in \cite{BFF01}:
\be
U^{\mu\nu}={\kappa\over 8\pi}\,\epsilon^{\mu\nu\rho} \,\eta_{ij}
\, B^i_\rho \,\left\{
\xi^j_{(V)}+\bar \xi^j_{(V)}\right\}
\label{22}
\ee
where $\Xi_P=\xi^\mu\partial_\mu +\xi^i\rho_i$
denotes now a generic vector field on the $SL(2,\Re)$
principal bundle  $P$ of the theory
and
\be
\xi^i_{(V)}= \xi^i+
 A^i_\mu \,
\xi^\mu\, ,\quad 
\bar \xi^i_{(V)}= \xi^i+
\bar A^i_\mu \,
\xi^\mu\label{23}
\ee
are the vertical parts of $\Xi_P$ with respect to $
A^i_\mu$ and $\bar A^i_\mu$, respectively. 

\noindent 
We recall in fact  that a vector field 
$\Xi_P$  on a principal bundle $P$ can be split, once a given connection
$A^i_\mu$ has been chosen, into the sum of its horizontal and vertical
parts:
\be
\Xi_P=\xi^\mu\partial^{(h)}_\mu +\xi^i_{(V)}\rho_i
\label{244}
\ee
where we have set:
\be
\cases{ \partial^{(h)}_\mu=\partial_\mu-A^i_\mu\,\rho_i \cr
\xi^i_{(V)}=\xi^i+  A^i_\mu \,
\xi^\mu}\  \label{254}
\ee
Notice that both the components $\xi^i_{(V)}$ and $\bar \xi^i_{(V)}$
in \form{23}  trasform  as vectors  under gauge transformations since
the non--tensorial character of $\xi^i$ is \emph{cured} by the
non--tensorial character of $A^i_\mu$.

Assuming that the topology of
spacetime is diffeomorphic to
$\Sigma\times \Re$ (see Appendix A), the Noether charge relative to the 
vector field $\Xi$ can be calculated from the
formula:
\be
Q_B(L_{\CCS},A,\bar A, \Xi)=\int_B U(L_{\CCS},,A,\bar A,
\Xi)\label{intsp}
\ee
where $B\simeq S^1$ is a $1$-dimensional surface embedded into $\Sigma$.
Now,  it would  seem physically reasonable to define the energy as the
Noether charge
relative 
to a spacetime vector field $\xi$ transverse to $\Sigma$, i.e. by
simply setting $\Xi=\xi^\mu\partial_\mu$ in   formula \form{22}, i.e.
$\xi^i=0$. Nevertheless, as it was already pointed out in
\cite{BFF01}, this prescription is not admissible from a
mathematical viewpoint since the group $\Diff(M)$ of spacetime
diffeomorphisms is not a global invariance group for the theory.
Indeed, in gauge natural theories the symmetry group  is the group
$\Aut(P)$  and $\Diff(M)$ is not a canonical and natural subgroup of it.
Namely, the splitting of $\Aut(P)$ into 
$\Diff(M)$ and $\Gauge(G)$ is   by no means  canonical and it is meaningful
only locally (we shall enter into the details of the matter below).
Roughly speaking, this means that, given a spacetime vector field
$\xi$, there is no canonical way to define a vector field $\Xi_P$
on the principal bundle and consequently to define the associated  vector
field
$\Xi$ on the configuration bundle
$Y$ which enters   \form{intsp}.
However, we can define, in a non canonical but nevertheless global
way,  a lift of spacetime vector fields through a dynamical
connection\footnote{The best known of these ways is the
\emph{horizontal} lift which consists simply in setting
$\xi^i_{(V)}=0$ in formula \form{244}.}. 
In the theory under examination
there are three different dynamical connections: the original
connections $A^i_\mu$ and 
$\bar A^i_\mu$ and their combination 
$\omega^i_\mu=( A^i_\mu+\bar A^i_\mu)/2$; accordingly,   different lifts
can be defined. In
\cite{BFF01}   the horizontal  lift with respect to  the
connection
$\bar A$  was considered. 
These choices correspond to set in
\form{22}: 
\be
\bar \xi^j_{(V)}=0,\qquad  \xi^j=- \bar A^j_\mu \,
\xi^\mu\label{24}
\ee 
 thus
leading to the Noether charge:
\be
Q_B(L_{\CCS}, \Xi)=\int_B {\kappa\over 8\pi}
 \, \eta_{ij} \,
B^i_\rho \, B^j_\mu \, \xi^\mu \, dx^\rho\label{intsp2}
\ee
(other choices will be considered in detail hereafter).
 We can now check the viability of  formula \form{intsp2}  by
specifying it for the BTZ solution.

In the $\mathfrak sl(2,\Re)$ basis \form{base} and in the coordinate
system $(x^\mu)=(t,\rho,\varphi)$ on spacetime, the connections $A=A^i_\mu
\,dx^\mu J_i$ and $\bar A=\bar A^i_\mu \,dx^\mu
J_i$ corresponding to the (exterior of) the BTZ black hole solution are
given by \cite{BH3,Mann93,Ca}
\be
{}^{(\pm)}A^i_\mu=\left(\begin{array}{ccc}
\pm {{r_+\pm r_-}\over{l^2}} \sinh\rho&0&{{r_+\pm r_-}\over{l}}
\sinh\rho\\
0&\pm 1& 0\\
{{r_+\pm r_-}\over{l^2}}
\cosh\rho&0&\pm {{r_+\pm r_-}\over{l}} \cosh\rho
\end{array}\right)
\label{solutionBTZ}
\ee
where the $(+)$ is referred to $A$ while $(-)$ is referred to $\bar A$.
For the triad field $e^i_\mu$ we have:
\be
e^i_\mu={l\over 2}\left( {}^{(+)}A^i_\mu-
{}^{(-)}A^i_\mu\right)=\left(\begin{array}{ccc}
{{r_+}\over{l}} \sinh\rho&0&{{ r_-}}
\sinh\rho\\
0&l& 0\\
{{ r_-}\over{l}}
\cosh\rho&0& {{r_+}} \cosh\rho
\end{array}\right)\label{18}
\ee
where
\be
M={r_+^2+r_-^2\over 8G l^2},\qquad J={r_+ r_-\over 4G l}
\ee
are, respectively, the mass and the angular momentum of the black
hole. The BTZ metric $g$ of
components $g_{\mu\nu}=\eta_{ij}\,e^i_\mu\,e^j_\nu$ then becomes:
\be
g=-\sinh^2 \rho (r_+ /l \,dt -r_- d\phi)^2 +l^2 d\rho^2 +\cosh^2 \rho
(r_-/l\, dt -r_+ d\phi)^2\label{gbtz}
\ee
Defining the surface $B$   as the surface of
constant $t$ and constant $\rho$, 
from formula \form{intsp2}
and the solution
\form{solutionBTZ} we obtain:
\be
\begin{array}{rl}
Q_B(L_{\CCS}, {}^{(\pm)}A, \hat{\partial_t})&={J\over l}\\
\\
Q_B(L_{\CCS}, {}^{(\pm)}A, \hat{\partial_\phi})&={1\over 4Gl}\left(\cosh^2
\rho\, r_+^2- \sinh^2 r_-^2
\right) 
\end{array}\label{202}
\ee
having denoted with 
 $\hat{\partial_t}$ and $\hat{\partial_\phi}$, 
respectively, the lift of the spacetime vector fields $\partial_t$ and
$\partial_\phi$ through the prescription \form{24}. These results do
not agree with 
the expected 
physical quantities
$M$ and
$J$, respectively, even in the limit $\rho\ram \infty$; see
\cite{btz,Mann94,BH,Mann93}. This can be seen as an hint that the
superpotential or the lift chosen in
\cite{BFF01}, or both,  are not suited in calculating physical
observables, at least in the Chern--Simons formulation of
$(2+1)$ gravity.
\CVD
}
\end{Example}
\vspace{1.2truecm}
Let us then consider again the expression \form{corr} and let us 
perform  the variation of the Noether current $\c{E}  ( L, \varphi,
\Xi )$ with a (vertical) vector field $X=\delta \varphi
{\partial\over\partial
\varphi}$. We obtain:
\ba
\delta_X {\c{E}}  (L, \varphi, \Xi)
&=&\delta_X<\Bbb{F} ( L, j^1\varphi),
\pounds_{{\Xi}}
\varphi> - i_\xi [\delta_X{L} (j^1\varphi)]\nonumber
\\
&=&\delta_X <\Bbb{F} (  L, j^1\varphi), \pounds_{{\Xi}} \varphi
>-\pounds_\xi <\Bbb{F} ( L, 
j^1\varphi ),X>\nonumber\\
&&-i_\xi <\Bbb{E} (L, j^2\varphi ),X>+d [  i_\xi <\Bbb{F} (
L, 
j^1\varphi ),X >] \nonumber\\
&=&\omega ( \varphi, X, \pounds_{{\Xi}} \varphi )+ d  (i_\xi <\Bbb{F}
(L, j^1\varphi) ,X>)\nonumber \\
&&- i_\xi <\Bbb{E} (L,j^2 \varphi ),X>
\label{trev}
\ea
where, in passing from the first to the second equality,  we made use of
\form{dL} and of the rule $\pounds_\xi=d\,  i_\xi + i_\xi\,  d$.
In \form{trev} we have denoted with 
 $\omega ( \varphi , X, \pounds_{{\Xi}} \varphi)$  the naive
symplectic current \cite{Waldsymp,Wald}:
\be
  \omega  ( \varphi, X, \pounds_{{\Xi}} \varphi)=\delta_X <\Bbb{F}
(L,  j^1\varphi),\pounds_{{\Xi}} \varphi>-\pounds_\xi  <\Bbb{F} (L,  
j^1\varphi  
), X > \label{nsycu}
\ee
 For field theories described through  a first order
Lagrangian  the Poincar\'e--Cartan morphism $<\Bbb{F} (  L,
j^1\varphi),X>$
 depends on the fields
$\varphi$ together with their first derivatives and depends linearly on the
components of the vector field $X$ (see \form{166}). Hence formula
\form{nsycu}, by using Leibniz rule,  can be rewritten:
\ba
  \omega  ( \varphi, X, \pounds_{{\Xi}}\varphi)&=&<\delta_X\Bbb{F}
(L,  j^1\varphi),\pounds_{{\Xi}} \varphi>-  <\pounds_\Xi\Bbb{F}  
(L,  j^1\varphi), X >\nonumber\\ &&+
<\Bbb{F} (L,  j^1\varphi),\delta_X (\pounds_{{\Xi}}
\varphi)-\pounds_\Xi X >
\label{nsycu2}
\ea
Moreover, for first order theories, the Poincar\'e--Cartan morphism,
which in practice is given by the derivatives of the Lagrangian with
respect to the first derivatives of the fields, is nothing but the
mathematical object describing the generalized momenta conjugated to
the dynamical fields (see \cite{CADM}). In trying to establish a
correspondence with Classical Mechanics we could say that the first
two terms  in
\form{nsycu2} correspond to the expression $ \delta p\,\dot q- \dot
p\,
\delta q$ which generates the Hamilton equations of motion through
the rule
$\delta H=\dot q \delta p- \dot p \delta q$ (in this analogy the
time derivative of Classical Mechanics is replaced, in field
theories,  by the Lie derivative $\pounds_\Xi$). \\
What about the further contribution to 
\form{nsycu2}? Carrying  on  the analogy with Classical Mechanics
this term would correspond  to the expression
$
p\,[\delta
(d_t q)-  d_t{(\delta q)}]
$ which is clearly
equal to zero in Classical Mechanics since the time derivative commute
with the variation of the configuration variables. The situation is quite
different in field theories. To realize this property let us consider
gauge theories (i.e. theories where the dynamical field is the gauge
potential
$A^i_\mu$). In this case we have:
\be
\begin{array}{rcl}
\delta_X (\pounds_{{\Xi}} A^i_\mu) &=&
\delta_X\left\{\xi^\rho d_\rho
A^i_\mu+d_\mu \xi^\rho A^i_\rho + d_\mu \xi^i +C^i_{jh} A^j_\mu
\xi^h \right\}\\
\\
&=&\xi^\rho d_\rho
(\delta_X A^i_\mu)+d_\mu \xi^\rho(\delta_X A^i_\rho) + 
C^i_{jh} (\delta_X A^j_\mu) \xi^h\\
\\
&&+d_\mu (\delta_X \xi^i)+C^i_{jh}
A^j_\mu (\delta_X \xi^h)\\
\\
&=&\pounds_{{\Xi}}(\delta_X  A^i_\mu) + D_\mu (\delta_X \xi^i)
\end{array}
\ee
so that:
\be
\delta_X (\pounds_{{\Xi}} A^i_\mu)-\pounds_{{\Xi}}(\delta_X  A^i_\mu)=
D_\mu (\delta_X \xi^i)\label{39}
\ee
If we consider the case in which the components $\xi^i$ are built out
of  the dynamical fields (see, e.g. formula \form{24}) the variations 
$\delta_X \xi^i$ are different from zero and the term \form{39} does
not vanish. However, through an integration by parts procedure,
 the naive
symplectic current \form{nsycu2} for gauge theories  splits as:
\be
  \omega  ( \varphi, X, \pounds_\Xi \varphi)= \tilde{\omega}   (  \varphi, X, \pounds_\Xi \varphi)
+ d  [\tau  ( \varphi, X, \Xi) ]+ f(\Bbb{E} (L, \varphi, X)
\label{symp3}
\ee
where 
\be
\tilde{\omega}   (  \varphi, X,  \pounds_\Xi \varphi)=<\delta_X\Bbb{F}
(L,  j^1\varphi),\pounds_{{\Xi}} \varphi>-  <\pounds_\Xi\Bbb{F}  
(L,  j^1\varphi), X >\label{omegatilde}
\ee 
is the \emph{reduced} symplectic $(m-1)$--form
and $f(\Bbb{E} (L, \varphi, X))$ denotes a term proportional to the
equations of motion.\footnote{
We stress that a  splitting similar to
\form{symp3} occurs also  for higher order natural theories, e.g.
General Relativity, even though its origin is quite different (see
\cite{forth}). Indeed, the divergence term $\tau $ does not arise
from the third term of \form{nsycu2}, which is identically
vanishing for natural theories,   but from an integration by parts
applied to the first two contributions. In both cases, i.e. in the first
order gauge natural theories considered herein as well as in the  higher
order natural theories  treated in \cite{HIO,forth}, it   is the analogy
with Classic Mechanics which suggests  how to perform the splitting
\form{symp3}. All the terms in
\form{symp3} which are not in the form $ \delta p\,\dot q- \dot
p\,
\delta q$ have to be decomposed, through integrations by parts, under the
form
$d  [\tau  ( \varphi, X, \Xi) ]+ f(\Bbb{E} (L, \varphi, X))$.}

Collecting together formulae \form{trev} and \form{symp3} we have:
\be
\delta_X  {\c{E}} (L, \varphi, \Xi)=
\tilde{\omega}   (  \varphi, X,  \pounds_\Xi \varphi)
+ d  \left\{\tau  ( \varphi, X, \Xi) 
+i_\xi <\Bbb{F}
(L, j^1\varphi) ,X>\right\}
+\hbox{E.M.}\label{de1}
\ee
\\
where E.M means terms proportional to the  Euler--Lagrange morphism
$\Bbb{E} (L, j^2\varphi, X)$ and hence vanishing on--shell. On the other
hand,  from formula \form{corcon} we have 
\be
\delta_X \c{E}  ( L, \varphi, \Xi )=\delta_X \tilde{\c{E}}  ( L,
\varphi, \Xi)+ d [ \delta_X{U}  ( L, \varphi, \Xi)]  \label{de2}
\ee
Comparing \form{de1} with \form{de2} we finally obtain:
\be
\begin{array}{c}
\delta_X \tilde{\c{E}}  ( L, \varphi, \Xi)+ d [ \delta_X {U} ( L, 
\varphi, \Xi)-  i_\xi < \Bbb{F} ( L, \varphi),  X>-
\tau  (  \varphi , X, 
\Xi) ]=\\
\\
\phantom{dfgdgadgdf } = \tilde{\omega}   (  \varphi, X,
 \pounds_\Xi \varphi )+\hbox{E.M.}
\end{array}\label{de3}
\ee
This formula can be seen as the counterpart in field theories of the
variational equation $\delta H =\dot q\, \delta p -\dot p\delta q+
[d_t(\partial L/\partial \dot q) -\partial L/\partial q]\delta q$ of
Classical Mechanics. This analogy suggests to define the variation
$\delta_X \H$ of the Hamiltonian density conjugate to the vector
field $\Xi$ as follows:
\ba
&\delta_X &\!\!\!\!\!\left[\H( L, \varphi, \Xi)\right]=\label{deltaH}\\
&=&\delta_X \tilde{\c{E}}  ( L, \varphi,
\Xi)+ d [ \delta_X {U} ( L, 
\varphi, \Xi)-  i_\xi <\Bbb{F} ( L, \varphi),  \pounds_\Xi \varphi> -
\tau  ( \varphi , X, 
\Xi) ]\nonumber
\ea
so that 
\be
\delta_X \H( L, \varphi, \Xi)=\tilde{\omega}   ( \varphi, X,
 \pounds_\Xi \varphi) +\hbox{E.M.}\label{Hamiltone}
\ee
Given a Cauchy hypersurface  $\Sigma$ the variation $\delta_X H$ of the
Hamiltonian is simply defined as $\delta_X H=\int_\Sigma \delta_X \H$.

 We remark  that  the right hand side of the 
equation \form{de3} does not contain divergence terms at all. This
means  that the divergence terms $d [ \delta_X {U} -  i_\xi \Bbb{F} -
\tau   ]$
in
\form{de3}  exactly cancel out  the divergence terms arising in the
variation
$\delta_X
\tilde{\c{E}}  $, i.e.
\be
d\left( {\partial \tilde{\c{E}} \over \partial (d\varphi)}
\delta_X\varphi
\right)=-d ( \delta_X {U} -  i_\xi \Bbb{F}-
\tau   )\label{SILVA}
\ee
hence leading to a formula for $\delta\H$ which is
divergence--free and which gives rise to the proper Hamilton equations
of motion \form{Hamiltone}. 

The  definition \form{deltaH} of the variation of the
Hamiltonian density  is   close  in spirit with  (and it can be seen
as a covariant generalization of)  the  original idea  of   Regge and
Teitelboim to handle boundary terms: all boundary terms arising in
the variation $\delta H$ of the Hamiltonian are added (with a minus
sign)   into the definition of the naive Hamiltonian in order to define 
(the variation of) a new Hamiltonian function endowed with a well defined
variational principle and hence suited  to be used as the generator  of
the allowed surface deformations; see
\cite{forth,Silva,RT}.
 \\
The terms of \form{deltaH} under the exterior differential are the
only ones surviving on shell, i.e. when we consider  a field $\varphi$  which is a
solution of field equations (we remind that $\tilde{\c{E}}$  is
proportional to field equations) and a variation $\delta_X$ performed along the
space of solutions (i.e.
$\delta_X
\tilde{\c{E}} =0 $).
  The variation $\delta_X Q $ of the Noether charges  relative to a
particular solution $\varphi$, relative  to the surface $\Sigma$ and
relative to the vector field $\Xi$, are hence  defined as the 
integral on the boundary $\partial \Sigma$:
\be
\delta_X Q_{\partial\Sigma}  (L, \varphi, {{\Xi}} )=
\int_{\partial \Sigma}  \delta_X {U} ( L, \varphi, {{\Xi}})-  i_\xi 
<\Bbb{F} ( L, \varphi),X>-\tau  ( \varphi , X, \Xi) \label{varqq}
\ee
From  equation \form{Hamiltone} it then follows that $\delta_X Q$
satisfies, on--shell,  the master equation:
\be
\delta_X Q  (L, \varphi, {{\Xi}} )=\int_\Sigma\tilde{\omega}   ( \varphi, X,
\pounds_\Xi \varphi)\label{1111}
\ee
\begin{Remark}{\rm
Definition \form{varqq} and equation \form{1111} deserve now some further
comment. \\
{\bf 1--} First of all we stress that  the (variation of) conserved
quantities is not defined only through the (variation of) of the
superpotential since two more corrective terms have to be added in the
definition. This is one of the reasons why formula \form{202} leads
to a wrong result.\\

{\bf 2--} The definition \form{varqq} do not depend on the
representative $L$ chosen inside the
homology class
$[L]$ of Lagrangians (two Lagrangians $L$ and $L'$ belong to the same
class $[L]$ if they differ only for divergence terms, which entails that 
they give rise to the same equations of motion). This property descends
from \form{SILVA}: all the terms $(\delta_X {U} 
-  i_\xi 
\Bbb{F} -\tau )$ which constitute the \emph{density} of the the
variation of the charges can be obtained directly and altogether
from the reduced current $\tilde{\c{E}}$. Since $\tilde{\c{E}}$ is
basically a linear combination of field equations with coefficients
given by the vector field $\Xi$ (see e.g. \cite{Remarks}, \cite{Lagrange} for a rigorous
proof of this statement) the only mathematical data we need for
defining
$\delta_X Q$ are the equations of motion and the generator of
symmetries $\Xi$. Nevertheless, even though \form{SILVA} can be used,
in practice,  as an operative schema to   calculate  explicitly
$\delta_X Q$ (we point out that this is exactly the
approach followed by Julia and Silva in
\cite{Silva}), from a theoretical point of view,  formula \form{SILVA}
hides the symplectic informations which are instead manifest in
\form{1111}.\\

{\bf 3--} Let us then consider  formula \form{1111}.
Recalling definition \form{omegatilde} we see that, if $\Xi$ is a
Killing vector  for the solution, i.\ e.\ $\pounds_\Xi\,\varphi=0$, 
the reduced symplectic form $\tilde \omega$ does vanish. From \form{1111}
we  obtain then:
\be
\delta_X Q  (L, \varphi, {{\Xi}} )=\int_{\partial\Sigma}\delta_X \Q(L,
\varphi, {{\Xi}} )=0\qquad  (\delta_X \Q=\delta_X {U} 
-  i_\xi 
\Bbb{F} -\tau  )\label{449}
\ee 
This latter equation is nothing but the conservation law of conserved
quantities. Indeed, 
let us assume that a metric on spacetime can be built out of the dynamical
fields, let us consider a (local) foliation of the
$m$--dimensional spacetime $M$ into space $+$ time and let us consider
a timelike hypersurface $\B$, namely a world tube in $M$; referring
to the example \form{gbtz} $\B$ would correspond to a surface with
constant $\rho$. Let us then denote by $B_{t_1}$ and $B_{t_2}$,
respectively,  the
$(m-2)$--surfaces generated by the intersection of $\B$ with two
spacelike hypersurfaces $\Sigma_{t_1}$ and $\Sigma_{t_2}$ at constant
times $t_1$ and $t_2$. Since $B_{t_1}\cup B_{t_2}$ defines a
boundary in
$\B$, from
\form{449} we infer that:
\be
\int_{B_{t_1}}\delta_X \Q(L,
\varphi, {{\Xi}} )-\int_{B_{t_2}}\delta_X \Q(L,
\varphi, {{\Xi}} )=0
\ee
so that, if $\left. \pounds_\Xi \varphi\right \vert_\B=0$,then  $\delta_X Q$
is conserved in time.

On the other hand let us consider a portion $D$ of a  spacelike
hypersurface $\Sigma_{t_0}$ at a given time $t_0$ (for example a
generic surface of constant $t$ for the BTZ spacetime \form{gbtz}) and
let us suppose that the (oriented) boundary $\partial \Sigma$ is formed by
the disjoint union of two $(m-2)$ surfaces $S$ and $S'$, e.g. two circles
of constant $\rho$ in the spacetime the metric of which is defined by
\form{gbtz}.  Assuming that $\Xi$ is a Killing vector on $\Sigma$, we have
\be
\int_{S}\delta_X \Q(L,
\varphi, {{\Xi}} )-\int_{S'}\delta_X \Q(L,
\varphi, {{\Xi}} )=0\label{522}
\ee
Referring now to the black hole solution and denoting by $S$ and $S'$,
respectively, the spatial infinity  and the horizon,
 formula \form{522} explains why conserved quantities such as mass and
angular momentum, which are naively calculated at spatial infinity   $S$, 
are related to properties of the horizon  $S'$: this comes from the
homological properties of $\delta_X Q$.

{\bf 4--} Formula \form{varqq} has a drawback: in fact  it provides
only the variation $\delta_X Q$ of conserved quantities  and the conserved
quantities $Q$ are obtained only after a formal integration.
Nevertheless the integrability of \form{varqq}  is not a priori
assured. It depends on the boundary conditions $\left.\delta_X
\varphi\right\vert_{\partial \Sigma}$ we impose.
Starting from the same expression, different boundary conditions may
lead to different results  corresponding to different physical
interpretations of conserved quantities.
For example
the recipe  \form{varqq}  has been proven to give the 
expected values   for the quasilocal energy  for Einstein and 
Einstein-Maxwell theories once the  variational equation is solved
with Dirichlet boundary conditions: see   \cite{HIO} and  \cite{forth}.\CVD}
\end{Remark}
\vspace{1cm}

We want hereafter to generalize the above formalism to the case of 
Chern-Simons theories with applications to Chern-Simons
$\hbox{AdS}_3$ gravity and to  BTZ black holes.
We assume as a Lagrangian for the theory the covariant Lagrangian 
(\ref{LB}).  The  superpotential is given by \form{22} and \form{23}
while the Poincar\'e--Cartan morphism has been calculated
in \cite{BFF01} to be the following:
\be
<\Bbb{F} (L_{\CCS},  A, \bar A), X>={\kappa\over
8\pi} 
\{ \epsilon^{\mu\nu\rho} \, \eta_{ij} \,
B^j_\rho \, \delta_X ( A^i_\nu+ \bar{A}^i_\nu) \}ds_\mu\label{535}
\ee
Hence, from definition \form{nsycu2} and the property \form{39} we
obtain (compare with \form{symp3}):
\ba
\omega^\mu ( A, \bar{A}, X, \Xi)&=&\tilde \omega^\mu (
A, \bar{A}, X, \Xi)+d_\nu\left\{\tau^{\mu\nu}( A, \bar{A}, X,
\Xi)\right\} \nonumber \\
&-& {\kappa\over 8 \pi} \epsilon^{\mu\nu\rho} \, \eta_{ij} 
\delta_X \xi^i \{   F^j_{\nu\rho}+\bar{F}^j_{\rho\nu}  \} \label{533}
\ea
with:
\ba
\tilde \omega^\mu (
A, \bar{A}, X, \Xi)&=&{\kappa\over 8 \pi} \epsilon^{\mu\nu\rho} \,
\eta_{ij}  [ \delta_X  B^j_\rho \pounds_\Xi (A^i_\nu+\bar A^i_\nu )-
\pounds_\Xi B^j_\rho \delta_X (A^i_\nu+\bar A^i_\nu ) ]\nonumber
\\
&=&{\kappa\over 2 \pi l} \epsilon^{\mu\nu\rho} \, \eta_{ij}  [ \delta_X 
e^j_\rho \pounds_\Xi \omega^i_\nu-
\pounds_\Xi e^j_\rho \delta_X \omega^i_\nu ]\label{54prime}
\ea
and
\be
\tau^{\mu\nu}( A, \bar{A}, X, \Xi))={\kappa\over 4  \pi} 
\epsilon^{\mu\nu\rho} \, \eta_{ij} \, B^j_\rho \, \delta_X ( \xi^i) 
\label{5544}
\ee
We stress that  the third  term in the right hand side of
\form{533}  identically vanishes on shell, so  that \form{533}
 reproduces exactly the structure  \form{symp3}.  We also stress  that
the  reduced  symplectic form $\tilde \omega$ as given by \form{54prime}
provides us
the correct symplectic   structure for General Relativity
once we have identified
$e^j_\rho$   with the vielbein and $ \omega^i_\nu$  with the spin
connection, according to \form{trans}. Then  
$e^j_\rho$ and $ \omega^i_\nu$ can be recognized as
dynamical variables  conjugated to each other, i.e. they form a pair
of $(q,p)$ variables in the appropriate phase space.\\ 

Inserting \form{22}, \form{535} and \form{5544} into the definition
\form{varqq} we finally obtain:
\be
\delta_X Q  (L_{\CCS}, A, \bar A, \Xi)={\kappa\over 4 \pi}  \eta_{ij}    
\int_B [\xi^i_{(V)}  \delta_X A^j_\mu-  \bar{\xi}^i_{(V)}  \delta_X 
\bar{A}^j_\mu ] d x^\mu\label{MASTER}
\ee
with $\xi^i_{(V)}=\xi^i + A^i_\mu\, \xi^\mu$, $\bar \xi^i_{(V)}=\xi^i +
\bar A^i_\mu\,
\xi^\mu$. Notice  that the above definition for the variation of the
charges in Chern--Simons theory  is clearly    covariant as well as gauge
invariant (indeed  $\delta_X A$, $\delta_X \bar A$, ${\xi}_{(V)}$ and 
$\bar{\xi}_{(V)}$ are all gauge vectors).
 \\

\section{Generalized Kosmann Lift}
Once the explicit formula \form{MASTER} for the variation of charges has
been established we are faced with  another problem. Our goal is to
make use of the same formula \form{MASTER} in order to calculate
different physical quantities, such as energy and  angular momentum,
for the BTZ black hole. To do that we have to appropriately
choose a vector field $\Xi$ on the configuration bundle $Y$ the  projection
$\xi$ of which  on $M$ is the generator of time traslations and angular
rotations. As already remarked this is not an obvious choice. Indeed
the vector fields $\Xi$ projecting onto the same spacetime vector field
are far from being unique! Given a vector $\Xi$ all the vectors 
obtained from it through the addition of a generic  vertical
vector  give in fact rise to the same projected vector.\\
Hence the fundamental problem is to find a  physically reasonable
mathematical  rule to lift up to  the configuration bundle a given spacetime
vector field. While in natural theories there exists a preferred
rule, namely the \emph{natural} lift (see, e.g.\ \cite{Kolar}), this is
not at all the case in gauge natural theories.\\
We now try to reformulate the problem in
terms of local coordinates. In the gauge natural theory we are
analysing, i.e. the theory described by \form{11},  an  infinitesimal
generator of Lagrangian symmetries is any vector field
$\Xi$ of the kind 
\form{20}, which is functorially associated to a given  vector field:
\be
\Xi_P=\xi^\mu\partial_\mu+\xi^i\rho_i\label{5588}
\ee
 on the relevant
$SL(2,\Re)$ principal bundle of the theory.
Given a spacetime vector field $\xi=\xi^\mu\partial_\mu$  the problem
to define its  lift up to the configuration bundle corresponds to   
the  problem of defining a rule for constructing the components $\xi^i$
 in \form{5588} starting from the components $\xi^\mu$, the
dynamical fields $A^i_\mu$, $\bar A^i_\mu$ together with  their derivatives.
This rule must be mathematically well--defined in the sense that $
\xi^i_{(V)}$ and $\bar
\xi^i_{(V)}$ must transform as vectors under gauge transformations.
For example, this requirement forbids us to simply set $\xi^i=
0$  since this
choice in general is not globally defined. 

In the sequel we shall consider
some of the admissible lifts we are allowed to construct. One of them is 
known in geometrically oriented literature 
as the  generalized  Kosmann lift of vector fields. 
The Kosmann lift was defined for the first time in \cite{Lorenzo} in order
to establish  a connection  between the ad hoc definition of Lie
derivative of spinor fields given in \cite{Kosmann}  and the general
theory  of Lie derivatives on fiber bundles.
We shall not
enter here into the mathematical details the definition of the Kosmann lift
is based on. For this
issue we refer the interested reader to
\cite{Godina} where an exhaustive bibliography can also be found. We shall
only specialize to the present case the formalism there developed. To this
end we just  outline that the generalized  Kosmann lift  we are going to
define  takes its values in the Lie algebra $so(1,2)\simeq sl(2,\Re)$. The
(generalized) Kosmann lift  is then available in the theory we are
analysing since 
$SO(1,2)$ is a reductive Lie subgroup of $GL(3,\Re)$; see \cite{Godina}.

Given a spacetime vector field
$\xi=\xi^\mu\partial_\mu $ the Kosmann lift $K(\xi)$ of $\xi$ on the
principal 
$SL(2,\Re)$  bundle locally  reads as:
\ba
\Xi_P=K(\xi)&=&\xi^\mu\partial_\mu+\xi^i_{(K)}\, \rho_i\\
\xi^i_{(K)} & =&{1 \over 2} \eta^{ij} \, \epsilon_{jkl} \;
\xi^{[k}_h 
\eta^{l] h}\label{xxx}\\
\xi^{k}_h &= & e^k_\rho ( e^\mu_h d_\mu \xi^\rho-\xi^\mu   d_\mu 
e^\rho_h )  \label{kosmud}
\ea
We again point out that 
the above expression for $\xi^i_{(K)}$ is obtained via skew--symmetrization
of the components
$\xi^{k}_h$ and hence it takes values in  the Lie algebra $so(1,2)$.
Moreover  the components $\xi^{k}_h$, in their turn, are built out only
from the components
$\xi^\mu$ and their derivatives. They are 
independent on any dynamical 
connection, as it is clearly shown by \form{kosmud}. However using the spin 
connection induced by the frame $e^a_\mu$:
\be
\omega^{a}_{b\mu}= e^a_\alpha (\Sigma^\al{}_{\beta
\mu}-\Sigma_\beta{}^\alpha{}_\mu+ \Sigma_{\mu\beta}{}^\alpha  )e^\beta_b\,
,
\qquad \Sigma^\al{}_{\beta
\mu}=e^\alpha_c\,\partial_{[\beta }\,e^c_{\mu]}
\ee
the 
expression  for the Kosmann lift can be rewritten 
in the following way:
\be
\xi^{k}_h =  e^k_\rho ( e^\mu_h d_\mu \xi^\rho-\xi^\mu   
d_\mu e^\rho_h 
)=e^\mu_h  \stackrel{(\omega)}{\nabla}_\mu( e^k_\nu 
\xi^\nu)-\omega^{k}_{h \mu} \xi^\mu \label{komconn}
\ee
The latter formula shows explicitly that the components
$\xi^a_{(V)}=\xi^a_{(K)}+A^a_\mu\xi^\mu$ and
$\bar \xi^a_{(V)}=\xi^a_{(K)}+\bar A^a_\mu\xi^\mu$ correctly transform  as
gauge vectors. Indeed we have:
\be
\begin{array}{rcl}
\xi^a_{(V)}&=&{1 \over 2} \eta^{a b} \, \epsilon_{bcd} \left\{
e^{\mu d}  \stackrel{(\omega)}{\nabla}_\mu( e^c_\nu 
\xi^\nu)+(A^{cd}_\mu-\omega^{cd}_{ \mu}) \xi^\mu
\right\}\\
&=&{1 \over 2} \eta^{a b} \, \epsilon_{bcd} 
e^{\mu d}  \stackrel{(\omega)}{\nabla}_\mu( e^c_\nu 
\xi^\nu)+{1\over l}e^a_\mu\,\xi^\mu\\
\bar\xi^a_{(V)}&=&{1 \over 2} \eta^{a b} \, \epsilon_{bcd} \left\{
e^{\mu d}  \stackrel{(\omega)}{\nabla}_\mu( e^c_\nu 
\xi^\nu)+(\bar A^{cd}_\mu-\omega^{cd}_{ \mu}) \xi^\mu
\right\}\\
&=&{1 \over 2} \eta^{a b} \, \epsilon_{bcd} 
e^{\mu d}  \stackrel{(\omega)}{\nabla}_\mu( e^c_\nu 
\xi^\nu)-{1\over l}e^a_\mu\,\xi^\mu
\end{array}\label{644}
\ee
The relations \form{xxx} and  \form{komconn} between the infinitesimal
symmetry generators
$\xi^\mu$ and 
$\xi^a_{(K)}$  suggest  how  different global lifts can be defined. Indeed,
in Chern-Simons theory with
$\mathfrak  g=sl(2,\Re) $ and three dynamical connections
$A$, $\bar  A$ and 
$\omega={{A + \bar A} \over 2}$, we have the possibility to introduce
 different  lifts which are mathematically well defined global lifts. 
 They can  be obtained formally from \form{komconn} by
 arbitrarily replacing one of  the dynamical connections  $A$, $\bar A$
and $\omega$ into the  expression for the covariant derivative and another
one into the last  term of  \form{komconn}.  For example we can choose the
lifts   defined by means of:
\be
\xi^{k}_h =e^\mu_h  \stackrel{(A)}{\nabla}_\mu( e^k_\nu 
\xi^\nu)-\omega^{k}_{h \mu} \xi^\mu 
 \label{liftwr}
\ee
where we have used the dynamical connections $A$ and $ \omega $.
Interchanging in this way the dynamical connections $A$, $\bar A$ and
$\omega$ we can define
 nine different lifts;
notice, however, that two of them, namely, the one with  $A,\bar A$ and
the one with $\bar A, A$, are both identical  to the Kosmann lift
\form{komconn} owing to the splitting \form{trans}. These lifts are global
as much as the Kosmann lift and there is no  mathematical prescription to
select one among them. However it can be easily shown  that each
different lift we can define   gives different values  for the Lie
derivatives of the dynamical fields and also different Noether
conserved quantities.  The choice we make among them   in  evaluating the
variation of global  charges (following the recipe  of formula
\form{MASTER}) has therefore to be  dictated by  pure physical
considerations. We then  proceed as follows. Evaluating the
different possible  lifts for  the two  spacetime vector fields which
generate, respectively,  time translations and angular rotations for  BTZ
solution 
\form{solutionBTZ}, we  construct the corresponding variations of
conserved quantities according to \form{MASTER} and we  look at the
ones reproducing the expected values of mass and angular momentum. That
seems to be  a physically  
 acceptable criterium to select among the different lifts. \\
For instance, performing the calculations using the Kosmann lift we obtain
the correct values for mass and angular momentum:
\ba
Q(L_{\CCS},K( \partial_t),B)&={r_+^2+r_-^2\over 8G 
l^2}=M\label{masaBTZ}\\
Q(L_{\CCS}, K(\partial_\phi),B)&={r_+ r_-\over 4G 
l}=J\label{mamaBTZ}
\ea
independently on the radius of the circle $B$ (to be rigorous, the
results \form{masaBTZ} and \form{mamaBTZ}
are correct modulo a constant of integration which can be viewed as the
charge of a background solution and can therefore be set equal to zero
fixing, in this way,  the zero level for the measurement of the charges).
For each one of the other six possible lifts we obtain  instead non
integrable expressions which, obviously, have no interest here.  Just to
show one let us  consider the lift of formula
\form{liftwr}; we obtain then  for the conserved quantity associated with
the lift of
$\xi=\partial_t$ the following:
\be
\delta Q={\kappa\over l^3}\left\{r_+\delta r_+ +r_-\delta r_- - r_-\cosh^2
(\rho)\, \delta r_+ +r_+\sinh^2 (\rho)\, \delta r_-\right\}
\ee
which is manifestly non integrable since 
${\delta\over \delta r_+}\left({\delta Q\over \delta
r_-}\right)\neq {\delta\over \delta r_-}\left({\delta Q\over \delta
r_+}\right)$. 

We stress that the only lift providing the expected results is then the
generalized  Kosmann lift. This  justifies a posteriori the choice of the
Kosmann lift to construct  the infinitesimal generator of symmetries for our
theory. We also stress again that, among all the possible lifts, the
Kosmann lift is the only one which does not involve a connection whatsoever
(see
\form{kosmud}). In a practical language we could say that, among the
different lifts, the generalized  Kosmann lift  is the ``most
natural''.\footnote{We point out that it is not natural, i.e. functorial,
in a rigorous mathematical language, since it does not preserve commutators:
\be
[K(\xi), K(\eta)]^{ab}=K([\xi,\eta])^{ab}+{1\over 2} e^{[a}_\mu
(\pounds_\eta g^{\mu\sigma}\pounds_\xi
g_{\sigma\nu})e^{b]\nu}
\ee
Only if at least one of the vector fields  $\xi$ or $\eta$ is a conformal 
Killing vector for the metric then
$[K(\xi), K(\eta)]=K([\xi,\eta])$. This is the reason why  the generalized
Kosmann lift  is sometimes referred to as  ``the quasi--natural'' lift in
gauge theories; see \cite{Godina}.}
\\ 

Let us now consider the results \form{masaBTZ} and \form{mamaBTZ}. As
we already pointed out these numerical values are independent on the radius
of the circle $B$ on which integration is performed. This fact is not
surprising. Indeed we know, see equation \form{1111} and expression
\form{54prime}, that, on shell, the master formula \form{MASTER} obeys the
equation:
\be
{\kappa\over 4 \pi}  \eta_{ij}    
\int_{\partial \Sigma} [\xi^i_{(V)}  \delta_X A^j_\mu-  \bar{\xi}^i_{(V)} 
\delta_X 
\bar{A}^j_\mu ] d x^\mu
=\int_\Sigma{\kappa\over 2 \pi l} \epsilon^{\mu\nu\rho} \, \eta_{ij}  [
\delta_X  e^j_\rho \pounds_\Xi \omega^i_\nu-
\pounds_\Xi e^j_\rho \delta_X \omega^i_\nu ]\label{700}
\ee
so that the left hand side is a homological invariant iff the right hand
side is vanishing. Let us then calculate the Lie derivatives $\pounds_\Xi
e^j_\rho$ and $\pounds_\Xi \omega^i_\nu$. From the general formula
$\pounds_\Xi A^i_\mu= \xi^\rho F^i_{\rho\mu}+D_\mu \xi^i_{(V)}$ and the
splitting \form{trans} we have:
\be
\begin{array}{rcl}
 \pounds_{\Xi }\, e^a_\mu&=&\xi^\rho d_\rho e^a_\mu+d_\mu\xi^\rho e^a_\rho+
\epsilon^a{}_{bc} e^b_\mu \xi^c\\
\\
\pounds_{\Xi }\,\omega^a_\mu&=& \xi^\rho d_\rho\omega^a_\mu+d_\mu\xi^\rho
\omega^a_\rho+ \stackrel{\omega}{D}_\mu {\xi^a}
\end{array}
\ee
When we take  $\Xi$ equal to  the (generalized)  Kosmann lift \form{kosmud}
we obtain:
\be
\begin{array}{rcl}
\pounds_{K(\xi)}\, e^a_\mu&=&{1\over 2} e^{\nu a} \pounds_\xi g_{\mu\nu}\\
\\
\pounds_{K(\xi)}\,\omega^i_{k\mu}&=&-{1\over 2}e^\nu_k e^{\rho i} D_\mu
(\pounds_\xi g_{\rho\nu}) +\pounds_\xi \Gamma^\rho_{\nu\mu} e^i_\rho
e^\nu_k
\end{array}\label{717}
\ee
where $\pounds_\xi$ denotes the usual Lie derivative with respect to the
spacetime vector field $\xi$. 
Notice that 
\be
\pounds_{K(\xi)} g_{\mu\nu}=\eta_{ij} \left\{ (\pounds_{K(\xi)}
e^i_\mu)\, e^j_\nu + e^i_\mu \, (\pounds_{K(\xi)}
e^j_\nu)\right\}=\pounds_\xi g_{\mu\nu}=\nabla_\mu \xi_\nu +\nabla_\nu
\xi_\mu
\ee
so that the generalized Kosmann lift 
reproduces  the usual Lie derivative of the metric with respect to
diffeomorphisms.
Both derivatives   \form{717} are
vanishing for the BTZ metric when $\xi$ is either $\partial_t$,
$\partial_\varphi$ or any linear combinations of them with constant
coefficients. In all these cases the right hand side of \form{700} is
vanishing so that:
\be
\delta_X Q(L_{\CCS},K( \partial_t+ \Omega
\partial_\varphi),B_1)=\delta_X Q(L_{\CCS},K(
\partial_t+ \Omega \partial_\varphi),B_2)\label{prova}
\ee
where $\Omega$ is a constant and $B_1$ and $B_2$ are two homologic
surfaces, i.e. they form the boundary $\partial\Sigma$ of a two dimensional
surface $\Sigma$.
\\

The above formula allows us to formulate   the first law of
black hole dynamics.
Indeed if we set $\Omega$ equal to the (constant)  angular velocity of the
BTZ black hole
(see \cite{BH3}):
\be
\Omega=\frac{r_-}{8 G l r_+}
\ee
and $B_1$ equal to any circle of constant radius enclosing the black hole
horizon  the left hand side of \form{prova} turns out to be:
\be
\begin{array}{rcl}
\delta Q(K( \partial_t+ \Omega
\partial_\varphi),B_1)&=&\delta Q( K( \partial_t),B_1)+\Omega
\delta Q( K(\partial_\varphi),B_1)\\
\\
&=& \delta M +
\Omega \delta J
\end{array}\label{first}
\ee
where  in the last equality we made use of the results \form{masaBTZ} and
\form{mamaBTZ}.
On the other hand  if in \form{prova} we take $B_2$ equal   to 
the outer horizon $H$ of the black hole (i.e. $\rho=0$) we have, after some
algebraic calculations:
\be
\delta Q(K( \partial_t+ \Omega
\partial_\varphi),H)= T \delta S\label{second}
\ee
where $T= \frac{(r^2_+-r^2_-)}{2 \pi l^2 r_+}$ is the temperature for
the BTZ black hole and $S=\frac{(2 \pi r_+)}{4
 G}$ is one quarter of the horizon area. Equating \form{first} and
\form{second} we obtain that the first law of black holes mechanics
\be
T \delta S= \delta M + \Omega \delta J \label{pruim}
\ee
holds true also in the domain of   a Chern--Simons formulation of $(2+1)$
gravity.


\section{Transition to General Relativity Variables}
\label{General Relativity}
In this section we shall show that formula \form{MASTER}, once we
specialize it for  the  Kosmann lift and we make use of the splitting
\form{trans}, 
reproduces   exactly the formula for the variation of conserved quantities
in General Relativity found elsewhere, see \cite{forth}. 
The  notations and the basic formulae 
relative to the ADM foliation of spacetime  entering the calculations of 
this section are summarized in the Appendix A.

For the sake of clarity we recall  here
the formula \form{MASTER}:
\be
\delta_X Q_B (L_{\CCS}, A, \bar A, \Xi)={\kappa\over 4 \pi} 
   \int_B \eta_{ij}  [\xi^i_{(V)}  \delta A^j_\mu-  \bar{\xi}^i_{(V)} 
\delta \bar{A}^j_\mu ] d x^\mu,\quad \kappa={l\over 4G}
\ee
We now insert  the splitting \form{trans} for  the
connections
${A}$ and 
$
\bar{A}$ in the above expression , and we make use of the   expression
\form{644} for
${\xi}^i_{(V)}$ and $\bar{\xi}^i_{(V)}$   obtained via  the  Kosmann
lift. This yields
\be
\delta_X Q_B  (L_{\CCS}, A, \bar A,K(\xi))=  {\kappa \over 4 \pi l}    
\int_B \epsilon_{aij}   [e^{\nu j } e^i_\rho \nabla_\nu \xi^\rho
\delta e^a_\mu+ e^a_\nu \xi^\nu \delta \omega^{ij}_\mu ] d x^\mu
\label{delqme}
\ee
where $
\omega^{a}_\mu=  {1 \over 2} \epsilon^a_{ij}    \omega^{ij}_\mu
$.
In the following we shall  calculate the  conserved quantities
\form{delqme}  with respect to the generalized Kosmann lift of  a
spacetime  vector field $\xi^\mu=N u^\mu+N^\mu$ (see Appendix \ref{appe}
for  the notation) and  our goal will be  to express the above formula in
terms of  metric quantities adapted to an orthogonal foliation  of
spacetime.

First of all we remind that, in the notations summarized in the Appendix
\ref{appe}, we have:
\be
\int_B\sqrt{g} f_\mu d
x^\mu =\int_B f_\mu\epsilon^{\mu
\nu
\rho} n_\nu u_\rho \sqrt{\sigma}
\ee
for any $1$--form $f=\sqrt{g} f_\mu dx^\mu$, being
$\sigma=\det(\sigma_{\mu\nu})$. By making use of the properties
$\epsilon_{aij} e^i_\mu e^j_\nu=\det(e)\,\epsilon_{Ô\rho\mu\nu} e^\rho_a$, 
$\epsilon^{\mu\nu\rho}\epsilon_{\alpha\beta
\gamma}=\delta^{\mu\nu\rho}_{\alpha\beta \gamma}$, taking into account
\form{une},  \form{vun} and the relations $e^\lambda_a\delta
e^a_\lambda=\delta \sqrt{g}/\sqrt{g}$, $\sqrt{g}= NV \sqrt{\sigma}$ (see
Appendix A) the
first term under integration in
\form{delqme} can be expressed as:
\be
\int_B \epsilon_{aij}   e^{\nu j } e^i_\rho \nabla_\nu \xi^\rho \delta
e^a_\mu d x^\mu=
\int_B \nabla^\mu \xi^\nu (u_\mu n_\nu-u_\nu n_\mu ) 
 \;\delta  \sqrt{\sigma} 
\ee
If now we apply Leibniz rule:
\ba 
&&u_\nu\nabla_\mu \xi^\nu=\nabla_\mu (u_\nu \xi^\nu)-\xi^\nu\nabla_\mu
u_\nu=-\nabla_\mu N-\xi^\nu\nabla_\mu
u_\nu\\
&&n_\nu\nabla_\mu \xi^\nu=\nabla_\mu (n_\nu \xi^\nu)-\xi^\nu\nabla_\mu
n_\nu=-\xi^\nu\nabla_\nu
n_\nu
\ea
and we recall formulae
\form{KK} and \form{HH} we get:
\ba
\int_B  \epsilon_{aij}   e^{\nu j } e^i_\rho \nabla_\nu \xi^\rho \delta
e^a_\mu d x^\mu&=& \nonumber  \int_B \delta \sqrt{\sigma} [u^\mu 
\xi^\nu
\Theta_{\mu \nu}+ n^\mu \nabla_\mu N- n^\mu  \xi^\nu  K_{\mu \nu}   ]=\\
&=&\int_B \delta
\sqrt{\sigma}[2 N (a^\mu n_\mu) -2 N^\mu  n^\nu  K_{\mu \nu}]   \label{t2}
\ea
where we  made use of $h^\mu_\alpha a_\mu=
h^\mu_\alpha\nabla_\mu N/ N$.

Calculations are more involved for the second term of  \form{delqme}.
We start from the compatibility condition:
\be
D_\gamma e^i_\mu=\nabla_\gamma e^i_\mu+ \omega^i_{j\gamma} e^j_\mu=0
\quad (\nabla_\gamma e^i_\mu=d_\gamma
e^i_\mu-\Gamma^\rho_{\mu\gamma}e^i_\rho)
\ee
to express the spin connection as
$\omega^i_{j \mu}=- e^\nu_j \nabla_\mu e^i_\nu$.
After some algebraic calculations we then
 obtain:
\ba
\int_B \epsilon_{aij}    e^a_\nu \xi^\nu \delta \omega^{ij}_\mu d
x^\mu&=&\int_B
\xi^\nu \delta^{\gamma \alpha \beta}_{\nu \mu
\sigma} u_\alpha n_\beta e^\mu_i\left[e^\sigma_j \delta e^j_\rho 
\,\nabla_\gamma e^{i
\rho}+ \delta(\nabla_\gamma e^{i
\sigma}) \right]
\sqrt{\sigma}\nonumber \\
 &=&2\int_B \sqrt{\sigma} [\xi^\gamma
\delta (n_\beta \nabla_\gamma u^\beta)- N  \delta
 (\nabla_\gamma n^\gamma)+N e^{i \beta} \delta e^\gamma_i
\nabla_\gamma n_\beta ] \nonumber
\ea
Since  $e_{i (\beta} \delta e_{\gamma)}^i=2 \delta g_{\beta\gamma}$,
using formulae \form{94} together with \form{ktun} and \form{HH}, we can
rewrite:
\ba
 &&\int_B\epsilon_{aij}    e^a_\nu \xi^\nu \delta \omega^{ij}_\mu d
x^\mu= 2\int_B \sqrt{\sigma} \{  N \delta \c{K}- N^\alpha \delta
(n_\beta K^\beta_\alpha) + {N\over 2}K^{\alpha \beta} \delta \sigma_ {\alpha
\beta}     \}
\label{t1}\\
 &&\phantom{\int_B}= \int_B\left\{2 N  \delta ( 
\sqrt{\sigma} \c{K}) -2
\sqrt{\sigma}  N^\alpha \delta (n_\beta K^\beta_\alpha)-N \sqrt{\sigma}
( \c{K}
\sigma^{\alpha \beta}-  \c{K}^{\alpha \beta} )
\delta \sigma_{\alpha \beta}\right\} \nonumber 
\ea
Summing up the two terms   \form{t2} and \form{t1} and multiplying the
result with $\kappa/4\pi l=1/ 16\pi G$,
 we finally obtain the explicit
formula for the variation of the conserved quantity relative to the
Kosmann lift of the vector field $\xi$.
It reads:
\be
\delta_X Q_B  (L_{\CCS}, A, \bar A, K(\xi))
= \int_B \left\{ N  \delta_X \c{E}- N^\alpha \delta_X J_\alpha+N {
\sqrt{\sigma}
\over 2} s^{\alpha \beta} 
\delta_X \sigma_{\alpha \beta}  \right\}\label{deltaqrg}
\ee
where:
\be
\cases{
 \c{E}=  {1 \over 8 \pi G}  \sqrt{\sigma}   \c{K}\cr
\cr
{J}_\alpha=   \frac{1}{ 8 \pi G }  \sqrt{\sigma} \sigma_\alpha^\mu 
{K}^\nu_\mu  {n}_{\nu} \cr
\cr
{s}^{\mu \nu}=\frac{1}{8 \pi G}  [({n}^{\alpha} {a}_{\alpha}) \sigma^{\mu
\nu}-\c{K} \sigma^{\mu \nu}+\c{K}^{\mu \nu} ]
}
\ee
are, respectively, the quasilocal energy density, the quasilocal angular
momentum density and the surface pressure; see \cite{Mann94,BY}.

 Formula
\form{deltaqrg}   reproduces exactly the formula for the  variation of 
conserved quantities   found in
\cite{Booth,Mann94,BY,forth,Kij}  for General Relativity. We  stress again
that the full correspondence between the variation of conserved quantities
in the
$SL(2,\Re)$ Chern--Simons (covariant)
theory and $(2+1)$ General Relativity essentially depends on the choice of
the generalized Kosmann lift, thereby selecting the Kosmann lift, whenever
it can be defined (see \cite{Godina}),  as the preferred lift to be
considered in the domain of gauge natural theories. This is also in
accordance with the results of
\cite{vangoden,Matteucci}  where the Kosmann lift was used to calculate the 
superpotential in the tetrad--affine formulation of General Relativity,  
as well as  with the results of 
\cite{BTZ} where the same lift was considered in the domain of BCEA
theories. Nevertheless we again point out that the generalized Kosmann
lift is just one among the various possibilities and there does not exist a
mathematical reason to select it, while  there exists only, a posteriori, 
 a physical justification for its choice;
see \cite{Matteucci}.


\section{Acknowledgments}
  We are grateful to A.\ Borowiec of the  University of Warsaw and to 
L.\ Fatibene, M.\ Ferraris and 
M.\  Godina of the
University of Torino for useful 
discussions and suggestions on the
subject. We mention that this research  has been performed under no
support  from the Italian Ministry of Research.


\begin{appendix}
\section{The Orthogonal Foliation of Spacetime\label{appe}}
We consider  a three dimensional region $D\subseteq M$ of a Lorentzian
three dimensional
manifold $(M,g)$ and a foliation of it into  spacelike
hypersurfaces 
$\{\Sigma_t\}$, being $t\in \Re$ the parameter of the foliation. Denoting
by 
$B_t$ the boundary  
$B_t=\partial
\Sigma_t$ of each leaf of the foliation the union $\B=\cup_{t\in \Re} B_t$
defines a timelike hypersurface, the unit normal of which we denote by
$n^\mu$.  We
shall restrict the attention solely to the case of  orthogonal foliations,
i.e. foliations for which it holds 
$\left. u^\mu n_\mu\right\vert_\B=0$, having 
 denoted by $u^\mu$ the components of the unit  vector field
which is everywhere orthogonal to each $\Sigma_t$ (see, e.g. \cite{BY,BLY}).

On the hypersurface $\B$ the spacetime metric $g$  can be
decomposed as:
\be
g_{\mu \nu}= \sigma_{\mu \nu} + {n}_\mu {n}_\nu - {u}_\mu {u}_\nu
\ee
where $ \sigma_{\mu \nu}$ is the metric induced by $g_{\mu \nu}$ on the
surface
$B_t$.  The metric can be also expressed in terms of the triad fields
$e^i_\mu$ as
$g_{\mu\nu}=\eta_{ij} e^i_\mu e^j_\nu$ where $\eta_{ij}={\mathop{\rm
diag}\nolimits}(-1,1,1)$. In
the sequel, on  each surface $B_t$ we shall use the following notation:
\be \label{une}
\cases{
  {u}_\mu ={e}^0_\mu \cr
  {n}_\mu ={e}^1_\mu}
\ee
so that 
$g$ results to be:
$
g_{\mu \nu}= \sigma_{\mu \nu} + {e}^1_\mu {e}^1_\nu - {e}^0_\mu {e}^0_\nu
$. Obviously we have:
 $ {u}_\mu {u}^\mu  ={e}^0_\mu {e}^{0 \mu}=-1$,
${n}_\mu {n}^\mu  ={e}^1_\mu {e}^{1 \mu}=1$, ${u}_\lambda  
{e}^\lambda_a= \delta^0_a$ and ${n}_\lambda {e}^\lambda_a= \delta^1_a $. 

In a system of coordinates $(t,\rho,\varphi)$  adapted to the foliation
and for which $\B$ is a constant $\rho$ hypersurface,   the  components
of the vectors
$u_\mu,n_\mu$ can be expressed as:
\be
\cases{
  {u}_\mu = (-N, 0,0) \cr
  {n}_\mu =(0,V,0)  }\label{deltanor}
\ee
where $N$ is the ordinary lapse in a foliation--adapted ADM decomposition
of the metric while
$V$ is called  the radial lapse. The vector field $\xi=\partial_t $  is
defined, in terms of the ADM lapse and shift,  as:
\ba
\xi^\mu= N u^\mu+N^\mu \label{tempo}
\ea
and the orthogonal condition $\left. u^\mu
n_\mu\right\vert_\B=0$  implies $\left. \xi^\mu
n_\mu\right\vert_\B=\left. N^\mu
n_\mu\right\vert_\B=0$. The variations $\delta_X g$ of the metric with
respect to a vertical  vector field $X=\delta g_{\mu\nu} {\partial\over
\partial g_{\mu\nu}}$ can be written as: 
\be
\delta_X g_{\mu \nu}= - {2 \over N}    {u}_\mu {u}_\nu  \delta N - {2
\over N}  h_{\alpha( \mu} u_{\nu)} \delta N^\alpha+ h^\alpha_{(\mu} 
h^\beta_{\nu)}\delta h_{\alpha\beta}\label{94}
\ee
where $h_{\mu \nu}= g_{\mu \nu} + {u}_\mu {u}_\nu$
are the components of the  metric on each $\Sigma_t$. Moreover, from
\form{deltanor}  we have:
\ba 
\delta  {u}_\mu ={\delta N \over N}  {u}_\mu  \label{vun}\\
  \delta {n}_\mu ={\delta V \over V}  {n}_\mu \nonumber
\ea
The extrinsic curvature of the generic  hypersurface $\Sigma_t$ embedded in
$M$ is  defined by:
\be
K_{\mu \nu}= -{h}_\mu^\alpha \nabla_\alpha u_\nu
\ee
while the extrinsic curvature of the hypersurface $\B$ embedded in $M$ 
results to be:
\be
{\Theta}_{\mu \nu}= -\gamma_\mu^\alpha \nabla_\alpha {n}_\nu
\ee
where $\gamma_{\mu \nu}= g_{\mu \nu} - {n}_\mu {n}_\nu$ is the metric on
$\B$ induced by $g$. The extrinsic  curvature of $B_t$ as a surface
embedded in
$\Sigma_t$ can be expressed, for each $t\in \Re$,  as:
\be
\c{K}_{\mu \nu}= -{\sigma}_\mu^\alpha D_\alpha n_\nu
\ee
where $D$ is the metric covariant derivative with respect to the metric 
$h$ of $\Sigma$.\\
The extrinsic curvatures of $\B$, $\Sigma$ and $B$ can be expressed each
one  in terms of the others via the relation \cite{BY,BLY}:
\be
{\Theta}_{\mu \nu}= \c{K}_{\mu \nu}+ (n^\alpha a_\alpha) u_{\mu} 
u_{\nu}+2 \sigma_{\alpha (\mu} u_{\nu)} K^{\alpha \beta} 
n_\beta\label{HH}
\ee
where $a$ is the covariant acceleration of the normal $u$, i.e. 
$a_\mu=u^\rho
\nabla_\rho  u_\mu$. Denoting by $b$ the covariant acceleration of the
normal
$n$, i.e. $b_\mu=n^\rho \nabla_\rho n_\mu$, we obtain the two formulae:
\be \label{ktun}
\cases{
\nabla_\rho u_\mu=- {K}_{\rho \mu }-u_\rho a_\mu   \cr
\nabla_\rho n_\mu=- {\Theta}_{\rho \mu }+n_\rho b_\mu
}\label{KK}
\ee

\section{ More Examples \label{generalsol}}
Apart from the BTZ solution, other solutions have been found for  
Chern-Simons field equations  in the framework  of Chern-Simons Gravity. We
are going to analyse the  Chern-Simons anti--de Sitter
 solution \cite{BH4}
and the single particle  solution  \cite{Deser,ma}. \\

\subsection{ The anti--de Sitter solution}

Let us  consider a three dimensional  manifold  $M$  with a boundary
$\partial M$ which has the topology of a torus.
 We assume $(\omega,\bar{\omega}, \rho)$\footnote{
These complex coordinates on spacetime are related to the usual spherical 
coordinates by means of the following expressions:
\be
\cases{
\omega=\varphi+ i t\cr
\bar{\omega}=\varphi- i t\cr
e^\rho =lr
}
\ee}
as coordinates over $M$ such that the boundary is located at 
$\rho =\infty$ and the torus is labelled by complex coordinates
$(\omega,\bar{\omega})$.
A solution 
of the $SL(2,\Co)$
Chern-Simons field equations  depending on   the coordinates on the torus 
can be found in \cite{BH4} and it  corresponds  to the  Euclidean  anti--de
Sitter solution. In the coordinates chosen a general
$SL(2,
\Bbb{C})$ connection reads as $A= A^i\, J_i$ with 
$A^i=A^i_\omega d \omega+ A^i_{\bar{\omega}} d \bar{\omega}+A^i_\rho d
\rho$.\footnote{
In agreement with the notation of \cite{BH4} the generators of $SL(2,\Co)$
are:
\be
J_1={i\over 2}\left(\begin{array}{ccc}
0&1\\
1&0
\end{array}\right),
\quad J_2=
{1\over 2}\left(\begin{array}{ccc}
0&-1\\
1&0
\end{array}\right),
\quad J_3={i\over 2}\left(\begin{array}{ccc}
1&0\\
0&-1
\end{array}\right)
\ee
}  Imposing the gauge condition:
\be
A_\rho=i J_3
\ee
and the boundary condition  $A_{\bar\omega}=0$ at $\partial M$, we obtain 
the following solution for the Chern-Simons equations $F_{\mu \nu}=0$:
\be \label{P1}
A^i_\mu=\left(\begin{array}{ccc}
\frac{-e^\rho l+4 e^{- \rho} L(\omega)G}{l}&0&0
\\
i \frac{e^\rho l+4 e^{- \rho} L(\omega)G}{l}&0& 0\\
0
&0&i
\end{array}\right)
\ee
in terms of an arbitrary  chiral function $L(\omega)$. With the same 
method we obtain the conjugate solution, by imposing $\bar
A_{\omega}=0$  at the boundary:
\be 
\bar{A}^i_\mu=\left(\begin{array}{ccc}
0&-\frac{e^\rho l-4 e^{- \rho} \bar{L}(\bar{\omega})G}{l}&0
\\
0&-i \frac{e^\rho l+4 e^{- \rho} \bar{L}(\bar{\omega})G}{l}&0\\
0
&0&-i
\end{array}\right)
\ee
The metric  $g_{\mu \nu}= \eta_{i j} e^i_\mu e^j_\nu$, where $e^i_\mu={l 
\over {2i}} (A^i_\mu-{\bar A}^i_\mu)$, turns out to be:
\be
ds^2=4 G l (L d \omega^2 +{\bar L} d { \bar \omega}^2)+( l^2 e^{2 \rho}+ 
16 G^2 L \bar L e^{-2 \rho}) d {\omega} d { \bar \omega}+ l^2 d
\rho^2\label{generalads}
\ee
and it is an exact solution  of Einstein equations  (with negative
cosmological constant $\Lambda=-1/l^2$) depending on two arbitrary 
functions
$L(\omega)$,
$\bar L(\bar \omega)$. 

We perform the calculations to obtain, from \form{MASTER} with
$\eta=\delta$ and $\kappa=-l/4G$, the variation
 of the conserved quantity
  relative  to the generalized $SO(3)$ Kosmann lift of the vector
field
$\xi=\partial_t=i
\Big(  {\pa
\over {\pa 
\omega}}-{\pa \over {\pa \bar{\omega}}}\Big)$. The variation of the
``mass'' results to  be:
\be
\delta_X Q_B(L_{\CCS},{K(\partial_t)},\; A, \; \bar A)={1\over
l}\left\{\delta_X  L(\omega)+\delta_X \bar{L}(\bar{\omega})\right\}
\ee
which can be integrated to obtain the ``mass'' for the anti--de Sitter
solution: 
\be
M:=Q_B(L_{\CCS},{K(\partial_t)},\; A, \; \bar
A)={1\over l}\left\{L(\omega)+\bar{L}(\bar{\omega})\right\}+
const\label{MMM}
\ee
With the generalized Kosmann lift  of the vector field 
$\xi=\partial_\phi= \Big( {\pa
\over {\pa 
\omega}}-{\pa \over {\pa \bar{\omega}}}\Big)$, we obtain:
\be
\delta_X Q_B(L_{\CCS},{K(\partial_\phi)},\; A, \; \bar A)=\delta_X 
L(\omega)-\delta_X \bar{L}(\bar{\omega})
\ee
such that the ``angular momentum'' results to be: 
\be
J:=Q_B(L_{\CCS},{K(\partial_\phi)},\; A, \; \bar
A)=L(\omega)-\bar{L}(\bar{\omega})+const\label{JJJ}
\ee
 We remark that we continued  to call the quantities \form{MMM} and
\form{JJJ}, respectively,  the ``mass'' and ``angular momentum''  since they
correspond exactly to these quantities  when the solution
\form{generalads}  becomes a black hole solution, i.e. when
$L=L_0$, $\bar L=\bar L_0$ and $Ml=L_0+\bar L_0$, $J=L_0-\bar L_0$.

\subsection{ The particle solution}

\noindent
In the same $\mathfrak sl(2,\Re)$ basis \form{base} and in the 
coordinates system $(t, \rho, \varphi)$ the exact solution
corresponding  to a   particle source in $(2+1)$ dimensional 
gravity turns out to be:
\be 
A^i_\mu=\left(\begin{array}{ccc}
\sqrt{ \rho^2+\gamma^2}&0&\sqrt{ \rho^2+\gamma^2}
\\
0&{1\over \sqrt{ \rho^2+\gamma^2}}& 0\\
\rho
&0&\rho
\end{array}\right)
\ee
\be  \label{P2}
\bar{A}^i_\mu=\left(\begin{array}{ccc}
-\sqrt{ \rho^2+\tilde\gamma^2}&0&\sqrt{ \rho^2+\tilde\gamma^2}
\\
0&-{1\over \sqrt{ \rho^2+\tilde\gamma^2}}& 0\\
\rho
&0&-\rho
\end{array}\right)
\ee
where $\gamma=1-\alpha$, $\tilde\gamma=1-\tilde \alpha$ and $\pi
(\alpha+\tilde \alpha)$ is the deficit angle of the conical
singularity introduced by the particle in the geometry of spacetime (see
e.g. \cite{Deser,Welling}).
\\ Using formula \form{MASTER} it is  possible to calculate the conserved 
quantities relative to the infinitesimal generators of symmetries in  
spacetime, suitably lifted by means of the  generalized Kosmann lift.
The  variations of mass and angular momentum turn out  to be:
\be
\begin{array}{rl}
\delta_X Q_B(L_{\CCS},{K(\partial_t)},\; A, \;\bar 
A)&=-{1\over l}(\gamma \delta_X  \gamma  +\tilde \gamma\; \delta_X  \tilde
\gamma)\\
\\
\delta_X Q_B(L_{\CCS}, {K(\partial_\phi)},\; A, \; \bar A)&=-(\gamma
\delta_X 
\gamma  -\tilde \gamma\;  \delta_X  \tilde \gamma)
\end{array}
\ee
Integrating the above expressions around the conical singularity we
obtain for the mass and the  angular momentum the following values:
\be
\begin{array}{rl}
 Q_B(L_{\CCS},{K(\partial_t)},\; A, \;\bar 
A)=-{1 
\over 2l} (\gamma^2 +\tilde\gamma^2)+ const={1\over l}\Big(L_0^+ + L_0^-
\Big)+ const
\\
\\
 Q_B(L_{\CCS},{K(\partial_\phi)},A, \; \bar A)=-{1 \over 2}
(\gamma^2 - \tilde\gamma^2) + const=L_0^+ - L_0^- +  const
\end{array}
\ee
where we have set $L_0^+=-{1 \over 2} (\gamma^2 )$ and $L_0^-=-{1 
\over 2} (\tilde\gamma^2)$.\\
These results  
 are in accordance with the values found by  
Martinec in \cite{ma} for  the ADM mass and spin of the particle.

\end{appendix}


\end{document}